\documentclass[10pt,aps,prd,preprintnumbers,superscriptaddress,nofootinbib,notitlepage,floatfix]{revtex4-2}
\usepackage[pdftex]{graphicx}
\usepackage{bm,latexsym,amsmath,amssymb,amsfonts,mathrsfs,physics}
\usepackage{acronym}
\usepackage{color}
\usepackage{empheq}
\allowdisplaybreaks[1]
\usepackage[pdftex,colorlinks=true,linkcolor=blue,citecolor=cyan,backref=page]{hyperref}
\newcommand*{\D}{\mathrm{d}}
\newcommand*{\mpl}{M_{\mathrm{Pl}}}
\newcommand*{\rmg}{\sqrt{-g}}
\newcommand*{\ns}{n_{s}}
\newcommand*{\ga}{g_{A}}
\newcommand*{\pol}{\sigma}
\newcommand*{\calg}{{\cal G}}
\newcommand*{\calf}{{\cal F}}

\newcommand*{\cald}{{\cal D}}
\newcommand*{\calp}{{\cal P}}
\newcommand*{\calr}{{\cal R}}
\newcommand*{\cali}{{\cal I}}

\newcommand*{\calq}{{\cal Q}}
\newcommand*{\cala}{{\cal A}}
\newcommand*{\calb}{{\cal B}}
\newcommand*{\dQ}{\delta Q}
\newcommand*{\ddQ}{\dot{\delta Q}}
\newcommand*{\dphi}{\delta\phi}
\newcommand*{\ddphi}{\delta\dot{\phi}}
\newcommand*{\dN}{\delta N}

\DeclareMathOperator{\diag}{diag}
\acrodef{CMB}{cosmic microwave background}
\acrodef{FLRW}{Friedmann-Lema\^itre-Robertson-Walker}
\acrodef{CNI}{chromo-natural inflation}
\acrodef{ADM}{Arnowitt-Deser-Misner}
\acrodef{EoM}{equation of motion}
\newacroplural{EoM}{equations of motion}
\acrodef{DoF}{degree of freedom}
\newacroplural{DoF}{degrees of freedom}
\acrodef{GW}{gravitational wave}
\acrodef{PGW}{primordial gravitational wave}

\begin{document}
\title{Chromo-natural inflation supported by enhanced friction from Horndeski gravity}
%
\author{Tomoaki~Murata}
\email[Email: ]{tmurata@rikkyo.ac.jp}
\affiliation{Department of Physics, Rikkyo University, Toshima, Tokyo 171-8501, Japan
}
\author{Tsutomu~Kobayashi}
\email[Email: ]{tsutomu@rikkyo.ac.jp}
\affiliation{Department of Physics, Rikkyo University, Toshima, Tokyo 171-8501, Japan
}
%
\begin{abstract}
We study the extension of the chromo-natural inflation model by incorporating nonminimal coupling between the axion field and gravity. 
Nonminimal coupling is introduced so that it enhances friction in the axion's equation of motion and thus supports slow-roll inflation.
This enhanced friction effectively delays the activation of the gauge field, thereby preventing the overproduction of gravitational waves in the CMB scale.
We extend previous results by describing the nonminimal coupling in a general and unifying way utilizing Horndeski gravity.
This allows us to explore systematically and comprehensively possible enhanced friction models of chromo-natural inflation consistent with observations.
We find a novel enhanced friction model that shows better agreement (within 1$\sigma$) with CMB measurements than the previous nonminimally coupled chromo-natural inflation model.
The gravitational-wave spectrum starts to rise at some wavenumber due to retarded activation of the gauge field in the late stage of inflation.
We show how one can identify the wavenumber at which this occurs based on the background evolution and present a universal analytic formula for the gravitational-wave spectrum that can be used for any enhanced friction model of chromo-natural inflation.
\end{abstract}
\preprint{RUP-24-11}
\maketitle
\section{Introduction}

It is strongly believed that the Universe underwent an accelerated expansion in its early stage, i.e. the period known as cosmic inflation. 
The accelerated expansion resolves several problems concerning the initial conditions of the Universe in the standard Big Bang theory such as the horizon and flatness problems~\cite{Starobinsky:1980te, Sato:1980yn, Guth_1981}.
Inflation also generates primordial curvature perturbations, which are considered to be the seeds for the large-scale structure observed in the present Universe. 
The nature of the inflationary perturbations can be seen in the statistical properties of the \ac{CMB} fluctuations, providing an observational link to the physics of inflation.
In addition to the curvature perturbations, inflation produces \acp{PGW}. 
These \acp{PGW} also carry valuable information about the physics of the inflationary universe. 
Although yet to be detected, confirming the presence of inflationary \acp{GW} by future observations would offer unparalleled insights into the early universe and hence is eagerly awaited.

If a gauge field existed during inflation, it could leave distinctive imprints on cosmological fluctuations. 
Among various models, axion inflation is motivated by particle physics and has attracted significant attention. 
In particular, an interaction between axions and U(1) gauge fields can produce notable fluctuations, which could be relevant to future observations~\cite{Anber:2009ua, Barnaby:2011vw, Sorbo:2011rz, Namba:2015gja, Peloso:2016gqs, Ozsoy:2021onx, Campeti:2022acx, Niu:2022fki, Fujita:2023inz, Caravano:2022epk, Figueroa:2023oxc, Caravano:2024xsb} (also, see~\cite{Maleknejad:2012fw, Komatsu:2022nvu} for recent reviews).
However, inflationary models with a U(1) gauge field are challenging because they would be incompatible with the isotropic background, which restricts the role of the U(1) gauge field in the inflationary dynamics to only a subdominant one.

Inflation involving interactions between the axion and SU(2) gauge fields has been proposed and named \ac{CNI}~\cite{Adshead:2012kp}.
The SU(2) gauge field is compatible with the isotropic cosmological background,
and it has been shown that the isotropic configuration is indeed an attractor solution~\cite{Maleknejad:2013npa, Wolfson:2020fqz, Wolfson:2021fya}.
The \ac{CNI} model could thus be a well-motivated, viable candidate for the inflationary scenario. 
Interestingly, the \ac{CNI} model generates large \acp{PGW} sourced by the tensor modes from the gauge field, which overwhelm those generated directly from vacuum fluctuations~\cite{Maleknejad:2016qjz, Papageorgiou:2018rfx}.
However, \acp{PGW} are rather overproduced and the original \ac{CNI} model has thus been ruled out by \ac{CMB} observations~\cite{Adshead:2013nka}.

A possible solution to the problem of the overproduction of \acp{GW} is using the axion as a spectator field rather than the inflation~\cite{Dimastrogiovanni_2017}.
The spectator \ac{CNI} model still possesses an interesting property of generating large \acp{PGW} that are within the current observational limits but could be detectable by next-generation experiments~\cite{Thorne:2017jft, Fujita:2018ndp, Campeti:2020xwn, LiteBIRD:2023zmo, Badger:2024ekb}.
The spectator \ac{CNI} model is the subject of intense research~\cite{Dimastrogiovanni:2012st, Mukohyama:2014gba, Maleknejad:2016qjz, Adshead:2016omu, Agrawal:2017awz, Caldwell:2017chz, Dimastrogiovanni:2018xnn, Agrawal:2018mrg, Fujita:2018vmv, Domcke:2018rvv, Lozanov:2018kpk, Papageorgiou:2019ecb, Mirzagholi:2020irt, Fujita:2021flu, Unal:2023srk, Fujita:2023axo}.
See also related works on inflation with non-Abelian gauge fields~\cite{Maleknejad:2011jw, Maleknejad:2011jr, Namba:2013kia, Nieto:2016gnp, Adshead:2017hnc, Iarygina:2021bxq, Fujita:2021eue, Fujita:2022fff, Murata:2022qzz}.

Another approach to the \ac{GW} overproduction problem is to explore modifications of coupling between the axion and gravity~\cite{Watanabe:2020ctz, Dimastrogiovanni:2023oid}. 
Recently, the derivative coupling of the axion to the Einstein tensor has been examined in the context of \ac{CNI} in~\cite{Dimastrogiovanni:2023oid}.
This nonminimal coupling enhances friction in the inflaton dynamics~\cite{Germani:2010gm}, delaying the activation of the gauge field and avoiding the overproduction of \acp{PGW} in the \ac{CMB} scale.

In this paper, we extend the previous work~\cite{Dimastrogiovanni:2023oid} to accommodate a more general coupling between the axion and gravity and investigate its implications on cosmological observations.
We do so by utilizing the general, unifying framework of enhanced friction models of inflation~\cite{Kamada:2010qe, Kamada:2012se, Mishima:2019vlh} developed in Horndeski gravity~\cite{Horndeski:1974wa, Deffayet:2011gz, Kobayashi:2011nu}.
(See~\cite{Kobayashi:2019hrl} for a review of Horndeski gravity.)
This approach allows us to explore wider regions of model space and parameter space of nonminimally coupled \ac{CNI} to look for models that better fit current observational data and lead to interesting predictions for future experiments.

The rest of this paper is organized as follows.
In the next section, we introduce a general description of enhanced friction models of \ac{CNI} in the Horndeski-SU(2) system and analyze the background slow-roll dynamics for several representative models of nonminimally coupled \ac{CNI}.
In Sec.~\ref{sec: CPT}, we overview the cosmological perturbations in the Horndeski-SU(2) system.
We then compute \ac{GW} spectra and \ac{CMB} observables and discuss their implications in Sec.~\ref{sec: Observation}.
Section~\ref{sec: conclusion} is devoted to conclusions.

\section{Inflation in Horndeski-SU(2) system}\label{sec: HorndekiSU(2)}

\subsection{Horndeski-SU(2) system}

Nonminimal couplings between axions and gravity have been investigated in the context of axion-SU(2) inflation in Refs~\cite{Dimastrogiovanni:2023oid, Watanabe:2020ctz}.
To study further generalizations of the nonminimal coupling,
it is convenient to use the Horndeski theory.
The action we consider is
\begin{align}
  S = S_{\rm H} + S_A.
  \label{action: HorndeskiSU(2)}
\end{align}
The Horndeski action $S_{\textrm{H}}$ is given by
\begin{align}
  S_{\rm H} &= \int\D^4x \rmg \biggl\{
  G_2(\phi, X)-G_3(\phi, X) \Box \phi
  +G_4(\phi, X) R+G_{4 X}\left[(\Box \phi)^2
  -\nabla_\mu \nabla_\nu \phi\nabla^\mu \nabla^\nu \phi\right]
  \notag \\ & \quad 
  +G_5(\phi, X) G_{\mu \nu} \nabla^\mu \nabla^\nu \phi
  -\frac{G_{5 X}}{6} \left[(\Box \phi)^3-3\Box \phi
  \nabla_\mu \nabla_\nu \phi\nabla^\mu \nabla^\nu \phi
  +2\nabla_\mu \nabla_\nu \phi\nabla^\nu \nabla^\lambda \phi\nabla_\lambda \nabla^\mu \phi
  \right]
  \biggr\},
\end{align}
where $X:=-\partial_{\mu}\phi \partial^{\mu}\phi /2$,
$R$ is the Ricci scalar, and $G_{\mu\nu}$ is the Einstein tensor.
The functions $G_2,\dots, G_5$ are in principle arbitrary,
but we will assume their forms more specifically below.
The action for the gauge field in \ac{CNI}, $S_{A}$, is given by
\begin{align}
  S_A = \int \D ^4 x \rmg 
  \qty(-\frac{1}{4} F_{\mu\nu}^a F^{\mu\nu}_a
  - \frac{\lambda}{4f} \phi
  F_{\mu\nu}^a \widetilde{F}^{\mu\nu}_a ),
\end{align}
where $\lambda$ and $f$ are model parameters,
$F^{a}_{\mu\nu}$ is a field strength of the SU(2) gauge field $A^{a}_{\mu}$,
\begin{align}
  F_{\mu\nu}^{a}
  = \partial_{\mu} A_{\nu}^{a}
  - \partial_{\nu} A_{\mu}^{a}
  + \ga {\epsilon^{a}}_{bc} A_{\mu}^{b} A_{\nu}^{c},
\end{align}
and $\widetilde{F}_{a}^{\mu\nu}$ is its dual,
\begin{align}
  \widetilde{F}_{a}^{\mu\nu}
  = \frac{1}{2\rmg}
  \epsilon^{\mu\nu\rho\sigma} F_{\rho\sigma}^{a}.
\end{align}
Here, $\ga$ is a gauge coupling constant, ${\epsilon^{a}}_{bc}$ is a structure constant of the SU(2) algebra, and $\epsilon^{\mu\nu\rho\sigma}$ is the totally antisymmetric symbol with $\epsilon^{0123}=1$.

It is clear that the above Horndeski-SU(2) system reduces to the standard one
for \ac{CNI} if one chooses
\begin{align}
    G_2=X-V(\phi),\quad V(\phi)=\mu^4\left(1+\cos\frac{\phi}{f}\right),
    \quad 
    G_4=\frac{\mpl^2}{2},
\end{align}
where $\mu$ is a constant and $\mpl$ is the Planck mass.
The nonminimally coupled model recently studied in~\cite{Dimastrogiovanni:2023oid}
in the context of \ac{CNI}
can be obtained by extending $G_4$ to\footnote{Note that the same nonminimal coupling can be expressed by using $G_5$ instead of $G_4$
as $G_5=-\phi/2M^2$.
However, Eq.~\eqref{eq:G4-ein-dpdp} is more convenient for generalizing the previous model.}
\begin{align}
    G_4=\frac{\mpl^2}{2}+\frac{X}{2M^2}\label{eq:G4-ein-dpdp}
\end{align}
and performing integration by parts to transform $XR$ to $G^{\mu\nu}\nabla_\mu\phi\nabla_\nu\phi$.
This nonminimal coupling enhances the friction term in the equation of motion for $\phi$ and thereby supports inflation~\cite{Germani:2010gm}.

In this paper, we generalize the previous work~\cite{Dimastrogiovanni:2023oid} and consider the functions of the form 
\begin{align}
  G_{2} = X - V(\phi) + M^4 F_{2}(\hat{X}),
  \quad
  G_{3} = M F_{3}(\hat{X}),
  \quad
  G_{4} = \frac{\mpl^2}{2} + M^2 F_{4}(\hat{X}),
  \quad
  G_{5} = \frac{1}{M} F_{5}(\hat{X}),\label{our-model}
\end{align}
where $F_2,\dots, F_5$ are dimensionless functions of 
\begin{align}
  \hat{X} := \frac{X}{M^4}.
\end{align}
Aside from the sinusoidal potential $V(\phi)$,
the Horndeski sector with Eq.~\eqref{our-model} respects the shift symmetry and thus
inherits the property of the axion.
As we will see below, the new additional terms $F_2,\dots, F_5$ provide generalized enhanced friction in the axion equation of motion.

\subsection{Inflationary background dynamics}

Let us investigate the cosmological background dynamics of the Horndeski-SU(2) system.
We consider the flat \ac{FLRW} universe,
\begin{align}
  \D s^2 = - N^2(t) \D t^2 + a^2(t) \delta_{ij} \D x^{i} \D x^{j},
  \label{eq: FLRWmetric}
\end{align}
where $N$ is the lapse function and $a$ is the scale factor.
We may impose $N=1$ by the coordinate choice.
The ansatz for the SU(2) gauge field that is compatible with the flat FLRW universe is given by
\begin{align}
  A^{0}_{i} = 0,
  \quad
  A^{a}_{i} = a (t) Q (t) \delta^{a}_{i}.
  \label{eq: GFansatz}
\end{align}
Using the FLRW metric and the ansatz for the gauge field,
one can derive the equations of motion for the metric, the axion field $\phi$, and the gauge field $Q$.%
\footnote{This ansatz is robust against initial anisotropies~\cite{Maleknejad:2013npa, Wolfson:2020fqz, Wolfson:2021fya} and initial spatial curvature~\cite{Murata:2021vnb}.}
The equations of motion for the cosmological background are presented in Appendix~\ref{App: Background} for shift-symmetric (but otherwise general) functions $G_3, G_4, G_5$.
For the functions given in Eq.~\eqref{our-model}, we obtain
\begin{align}
  3 \mpl^2 H^2
  &= X + V - M^4F_{2}
  + 2 X F_{2\hat{X}}
  + M^4 {\cal E}_{M}
  + \frac{3}{2} \qty(\dot{Q} + H Q )^2 + \frac{3}{2} \ga^2 Q^4,
  \label{eq:Friedmann00}
  \\
  - 2 \mpl^2 \dot{H}
  &=
  2 X \cali
  + 2 M^2 \dot{H} \qty[2 F_{4}
  - 2 \qty(\frac{\dot{\phi}}{M^2})^2 F_{4\hat{X}}
  - \qty(\frac{H}{M}) \qty(\frac{\dot{\phi}}{M^2})^3 F_{5\hat{X}} ]
  + 2 \qty(\dot{Q} + H Q )^2 + 2 \ga^2 Q^4
  \notag\\
  &\quad
  - M \ddot{\phi} \qty[ 2 \hat{X} F_{3\hat{X}}
  + 4 \qty(\frac{H}{M}) \qty(\frac{\dot{\phi}}{M^2}) F_{4\hat{X}}
  + 8 \qty(\frac{H}{M}) \qty(\frac{\dot{\phi}}{M^2}) \hat{X} F_{4\hat{X}\hat{X}}
  + 6 \qty(\frac{H}{M})^2 \hat{X} F_{5\hat{X}}
  + 4 \qty(\frac{H}{M})^2 \hat{X}^2 F_{5\hat{X}\hat{X}} ],
  \label{eq:Fr-ij}
\end{align}
where a subscript $\hat X$ denotes differentiation with respect to $\hat X$ and 
\begin{align}
  \cali &:= 
  1 + F_{2\hat{X}}
  + 3 \qty(\frac{H}{M}) \qty(\frac{\dot{\phi}}{M^2}) F_{3\hat{X}}
  + 6 \qty(\frac{H}{M})^2 F_{4\hat{X}}
  + 12 \qty(\frac{H}{M})^2 \hat{X} F_{4\hat{X}\hat{X}}
  \notag\\&\quad
  + 3 \qty(\frac{H}{M})^3 \qty(\frac{\dot{\phi}}{M^2}) F_{5\hat{X}}
  + 2 \qty(\frac{H}{M})^3 \qty(\frac{\dot{\phi}}{M^2}) \hat{X} F_{5\hat{X}\hat{X}},
  \label{eq: PhiConserved}
  \\ 
  {\cal E}_{M} &:=
  6 \qty(\frac{H}{M}) \qty(\frac{\dot{\phi}}{M^2}) \hat{X} F_{3\hat{X}}
  - 6 \qty(\frac{H}{M})^2 F_{4}
  + 24 \qty(\frac{H}{M})^2 \hat{X} \qty(F_{4\hat{X}} + \hat{X} F_{4\hat{X}\hat{X}} )
  \notag\\&\quad
  + 2 \qty(\frac{H}{M})^3 \qty(\frac{\dot{\phi}}{M^2}) \hat{X} \qty(5 F_{5\hat{X}} + 2 \hat{X} F_{5\hat{X}\hat{X}} ).
\end{align}
The equations of motion for the axion and the gauge field are given respectively by
\begin{align}
  &\cali\ddot{\phi}
  + 3 H \cali \qty(1 + \frac{\dot{\cali}}{3 H \cali} ) \dot{\phi}
  + V_{,\phi}
  = -\frac{3\ga \lambda}{f} Q^2 \qty(\dot{Q}+HQ ),
  \label{eq: EOMphi}
  \\ 
  &  \ddot{Q} + 3H\dot{Q} + (\dot{H} + 2H^2) Q + 2 \ga^2 Q^3
  - \frac{\ga \lambda}{f} Q^2 \dot{\phi} = 0.
\end{align}

Let us now consider the slow-roll inflationary dynamics of the above Horndeski-SU(2) system.
We assume that slow-roll inflation is supported by the gravitationally enhanced friction that arises if $\mathcal{I}\gg 1$ rather than the interaction between the inflaton and the gauge field.
Although this is true throughout the entire duration of inflation, the energy fraction of the gauge field gradually increases toward the end of inflation.
The friction-dominated inflationary dynamics aims to fit our model to observations,
given that the original \ac{CNI} model has already been ruled out by data.
As the energy fraction of the gauge field increases, its impact on the production of \acp{PGW} gets stronger, possibly yielding a signature detectable by future gravitational-wave experiments.

We assume the slow-roll conditions to hold:
\begin{align}
  \epsilon := - \frac{\dot{H}}{H^2}\ll 1,
  \quad
  \eta := \frac{\ddot{\phi}}{H \dot{\phi}}\ll 1.
\end{align}
It follows from these two conditions that 
\begin{align}
    \dot{\cali}\ll H\cali.
\end{align}
We assume in addition that
\begin{align}
    \frac{H}{M}\gg 1,
\end{align}
so that the friction term in the axion equation of motion is enhanced:
\begin{align}
    \cali\gg 1.
\end{align}
The effects of the gauge field are supposed to be subdominant
in the axion equation of motion,
\begin{align}
  \cali \dot{\phi}
  \gg
  \frac{\ga \lambda}{f} Q^3,
\end{align}
and thus, it is approximated as
\begin{align}
  3 \cali H \dot{\phi} + V_{,\phi} \simeq 0.
  \label{eq: SlowRollEOM}
\end{align}
This equation shows how the enhanced friction supports axion inflation without the help of the gauge field.

The gauge field may be considered as moving in the effective potential
\begin{align}
  U(Q)
  = H^2 Q^2 + \frac{1}{2} \ga^2 Q^4
  - \frac{\ga \lambda}{3f} \dot{\phi} Q^3.
\end{align}
The shape of $U$ depends on the value of $\dot\phi$, and it is easy to see that
the effective potential has a local minimum at $Q\neq 0$ if
\begin{align}
  \dot{\phi} 
  > \frac{4fH}{\lambda}
  \label{eq: ConditionOfLocal}
\end{align}
is satisfied.
The gauge field then tracks the slowly varying minimum:
\begin{align}
  \ga Q = \ga Q_{\rm min} := \frac{\lambda \dot{\phi}}{4f}
  + \sqrt{\qty(\frac{\lambda \dot{\phi}}{4f} )^2 -H^2}.
  \label{eq: QandPhi}
\end{align}
The amplitude of the gauge field $Q$ is thus controlled by the value of $\dot{\phi}$.

To see the background dynamics of inflation, it is convenient to introduce the density parameters representing a fraction of each term in the Friedmann equation~\eqref{eq:Friedmann00},
\begin{align}
  \Omega_{X} &= \frac{X}{3\mpl^2 H^2},
  \quad
  \Omega_{V} = \frac{V}{3\mpl^2 H^2},
  \quad
  \Omega_{M} = \frac{M^4{\cal E}_{M}}{3\mpl^2 H^2},
  \quad
  \Omega_{B} = \frac{\ga^2 Q^4}{2\mpl^2 H^2},
  \quad
  \Omega_{E} = \frac{\qty(\dot{Q} + HQ )^2}{2\mpl^2 H^2}.
  \label{eq: DensityParam}
\end{align}
Under our slow-roll approximation,
Eq.~\eqref{eq:Fr-ij} yields 
\begin{align}
    \mpl^2\dot H\simeq - X\cali
    \quad \Rightarrow\quad 
    \epsilon\simeq \frac{X\cali}{V}\gg \frac{X}{V},
    \label{eq: SlowRollEVOL}
\end{align}
and for $\cali\gg 1$ we have $X\cali\sim M^4\mathcal{E}_M$.
We, therefore, expect the hierarchy of the density parameters
$\Omega_{V} \gg \Omega_{M} \gg \Omega_{X},\Omega_{B},\Omega_{E}$ during inflation.
We will confirm this hierarchy with the numerical calculations in the following subsection.

\subsection{Numerical examples of the background evolution}\label{subsec:models}

So far, we have kept the functions $F_2,\dots, F_5$ arbitrary
and presented the general discussion on the slow-roll dynamics of the axion supported by the enhanced friction.
Now let us show some specific examples of the inflationary background evolution.
We switch on one of the functions $F_3, F_4, F_5$ and set it to be
$F_i=\hat X^n$ with $n$ being some number.
Then, we have
\begin{align}
    \cali\sim \left(\frac{H}{M}\right)^p\left(\frac{\dot\phi}{M^2}\right)^q,
    \label{eq: calI_param}
\end{align}
where the numbers $p$ and $q$ depend on which function we switch on and $n$.
(We do not consider $F_2$ because the term is minimally coupled to gravity and hence it does not yield an enhancement of $\cali$ by $H/M\gg 1$, i.e., $p=0$.)
More specifically, we consider the following three representative cases
with different pairs of $(p,q)$ to illustrate the background evolution:
\begin{align}
    \text{(i):}&\quad 
    F_3=\hat X\quad \Rightarrow\quad 
    \cali=3\left(\frac{H}{M}\right)\left(\frac{\dot\phi}{M^{2}}\right);
    \\
    \text{(ii):}&\quad 
    F_4=\hat X\quad \Rightarrow\quad 
    \cali=6\left(\frac{H}{M}\right)^2;
    \\
    \text{(iii):}&\quad 
    F_5=\hat X^{1/2}\quad \Rightarrow\quad 
    \cali=\sqrt{2}\left(\frac{H}{M}\right)^3.
\end{align}
The case (ii) reduces to the model studied in Ref.~\cite{Dimastrogiovanni:2023oid} with the rescaling $M\to\sqrt{2}M$.
Some more examples
(which turn out to be less interesting than the above three)
will be discussed in the next subsection.

\begin{figure}
   \begin{center}
     \includegraphics[keepaspectratio=true,height=42mm]{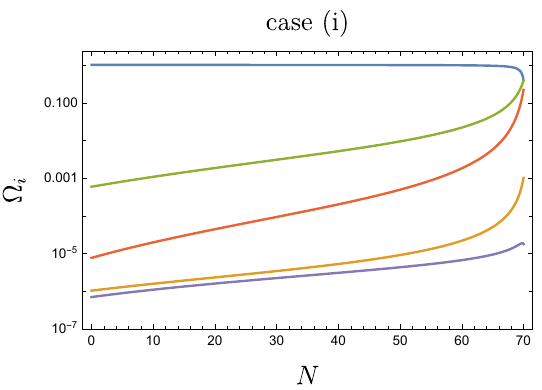}
     \includegraphics[keepaspectratio=true,height=42mm]{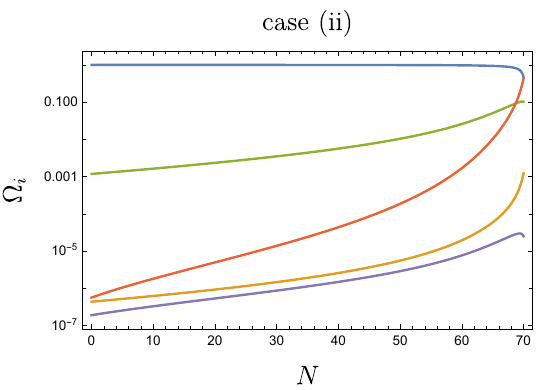}
     \includegraphics[keepaspectratio=true,height=42mm]{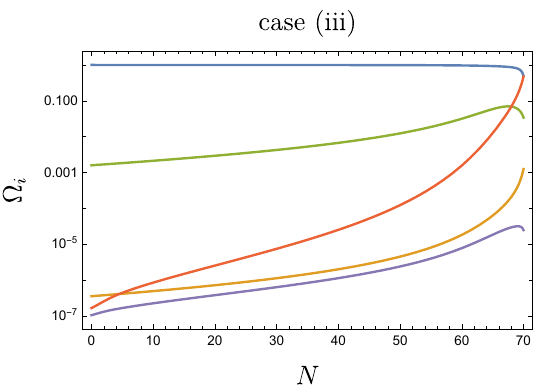}
  \end{center}
  \caption{Evolution of the density parameters~\eqref{eq: DensityParam} as functions of the e-folding number $N$ for the models with (i) $F_3=\hat X$, (ii) $F_4=\hat X$, and (iii) $F_5=\hat X^{1/2}$. 
  Parameters are given by Eqs.~\eqref{eq: FiducialParams} and~\eqref{eq:paras2}.
  Different lines correspond to $\Omega_{V}$ (blue), $\Omega_{X}$ (orange), $\Omega_{M}$ (green), $\Omega_{B}$ (red), and $\Omega_{E}$ (purple).
  }
  \label{fig: DensityParamBG}
\end{figure}

In Fig.~\ref{fig: DensityParamBG}, we show the background evolution of the models (i)--(iii) obtained numerically in terms of the density parameters defined in Eq.~\eqref{eq: DensityParam}.
In all these examples, the parameters are given by
\begin{align}
  \mu = 4.6 \times 10^{-3} \mpl ,
  \quad
  f = 0.2  \mpl ,
  \quad
  \ga = 0.05,
  \quad
  \lambda = 590.
  \label{eq: FiducialParams}
\end{align}
For each case, the parameter $M$ is taken to be 
\begin{align}
  \text{(i)}&\quad 
    1.98 \times 10^{-5} \mpl;
    \quad
    \text{(ii)}\quad 
    1.35 \times 10^{-6} \mpl;
    \quad
    \text{(iii)}\quad 
    1.77 \times 10^{-6} \mpl.\label{eq:paras2}
\end{align}
These values are chosen so that the amplitude of the curvature perturbation is consistent with CMB observations, as will be discussed in Sec~\ref{sec: Observation}.

We can see from Fig.~\ref{fig: DensityParamBG} that the different models yield qualitatively the same result.
In particular, the hierarchy of the density parameters derived based on our slow-roll assumptions is seen to hold except close to the end of inflation.
However, one can find an important quantitative difference when one focuses in detail on the relation between $\Omega_{B}$ and $\Omega_{X}$:
the time when $\Omega_{B}$ exceeds $\Omega_{X}$ depends on the model.
This can be understood by considering the magnitude of the gravitationally enhanced friction $\cali$.
Since $H/M\gg\dot{\phi}/M^2$, the gravitationally enhanced friction is more (less) effective for larger (smaller) $p$ and smaller (larger) $q$.
If the gravitationally enhanced friction is less effective, the impact of the gauge field appears earlier.
We will see in the next section how this difference crucially affects observational consequences of the enhanced friction models of \ac{CNI}.

\subsection{Other examples of the enhanced friction models} \label{sec: OtherFuncs}

Let us comment on the other choices of the Horndeski functions.
In addition to the models (i)--(iii),
we considered the models with (iv) $F_{3}=\hat{X}^{1/2}$
($p=1, q=0$) and (v) $F_{5}=\hat{X}$ ($p=3,q=1$).
As implied by the values of $p$ and $q$,
the effect of the enhanced friction turned out to be less effective in
these cases than in the model (iii).
Therefore, we do not present the detailed results of the models (iv) and (v).
We also considered hybrid models in which two of the Horndeski functions are switched on. However, we did not obtain physically new results compared to the models with a single nonvanishing Horndeski function.
Therefore, we also do not present a detailed analysis of hybrid models.
Note in passing that for $F_4= \hat X^{1/2}$, we have $\cali=0$, and there is no enhancement of the friction term.

\section{Cosmological perturbations} \label{sec: CPT}

In this section, we study the cosmological perturbations in the Horndeski-SU(2) system following and extending Refs.~\cite{Dimastrogiovanni:2012ew, Dimastrogiovanni:2023oid}.

We write the metric in the \ac{ADM} form as
\begin{align}
  \D s^2 =& -N^2 \D t^2 + \gamma_{ij}
  \qty(\D x^i+N^i \D t ) \qty( \D x^j+N^j \D t ).
\end{align}
In the spatially flat gauge, the lapse function, the shift vector, and the spatial metric are given respectively by
\begin{align}
  N = 1+\delta N ,
  \quad
  N_i = \partial_i B,
  \quad
  \gamma_{ij} = a^2 (e^{h})_{ij}
  = a^2 \qty(\delta_{ij} + h_{ij} + \frac{1}{2} h_{ik} {h^{k}}_{j} +\cdots ),
\end{align}
where $\delta N$ and $B$ are scalar perturbations and
$h_{ij}$ is a transverse traceless tensor perturbation.
After fixing the SU(2) gauge degrees of freedom, the perturbed gauge field is given by
\begin{align}
  A^{a}_{0} = a \partial_{a} Y,
  \quad
  A^{a}_{i} = a \qty[(Q+\delta Q) \delta_{ia}
  + \partial_{i} \partial_{a} Z + T_{ia}  ],
\end{align}
where $Y$, $\dQ$, and $Z$ are scalar perturbations and $T_{ia}$ is a symmetric, transverse and traceless tensor perturbation.
The scalar field is also perturbed as
\begin{align}
  \phi = \phi_{0} + \dphi.
\end{align}
Here, we used the subscript 0 to denote the background quantity,
but hereafter we will omit it to simplify the notation.
We do not consider vector perturbations, focusing on the power spectra of the curvature perturbation and \acp{GW}.

Substituting the perturbed quantities to the action and expanding it to second order in perturbations, we obtain the quadratic action for the perturbations.
Upon doing so in the scalar sector, we find that $\delta N$, $B$, and $Y$ are nondynamical variables whose equations of motion are constraint equations that can be solved to express them in terms of the dynamical variables $\delta Q$, $Z$, $\delta\phi$, and their derivatives.
Substituting the solutions to the constraint equations back to the action, we arrive at the quadratic action for the coupled system of the three dynamical variables $(\delta Q, Z,\delta\phi)$.
Although this is the legitimate way of dealing with the scalar perturbations, we notice that dropping the nondynamical variables in the metric perturbations, $\delta N$ and $B$, from the beginning and solving the constraint equation only for $Y$ still yields sufficiently accurate results until slightly after horizon crossing (see Appendix~\ref{app: MetricPtb}).
In this paper, we adopt this simplified treatment to reduce the computational cost.
The same simplified treatment was carried out in~\cite{Dimastrogiovanni:2023oid, Dimastrogiovanni:2012ew, Adshead:2013nka}.

In the tensor sector, we have two variables, $(h_{ij}, T_{ia})$, both of which are dynamical.
Thus, both scalar and tensor sectors of cosmological perturbations are described by coupled quantum systems.
Below we follow the quantization procedure given in Ref.~\cite{Nilles:2001fg}, a summary of which is presented in Appendix~\ref{App: Quantaization}.

\subsection{Tensor perturbations}

Let us derive the expression for the power spectrum of the primordial tensor perturbations from the Horndeski-SU(2) model.
We have two tensor variables $(h_{ij}, T_{ia})$,
and each has two polarization degrees of freedom.
Therefore, the Horndeski-SU(2) system has four degrees of freedom in the tensor sector.

We decompose the tensor perturbations into the circular polarization basis.
Performing a Fourier decomposition, we write
\begin{align}
  h_{ij} (t,\bm{x})
  = \int \frac{\D^3 k}{(2\pi)^3} e^{i\bm{k}\cdot\bm{x}}
  \qty[ h^{+}(t,{\bm{k}})e^{+}_{ij}(\bm{k})
  +  h^{-}(t,{\bm{k}})e^{-}_{ij}(\bm{k}) ],
\end{align}
and similarly for $T_{ij}$,
where $e^{\pm}_{ij}$ is the circular polarization tensors satisfying
\begin{align}
  i \epsilon_{ilm} k^{l} e^{\pm}_{jm}
  = \pm k e^{\pm}_{ij}.
\end{align}

After straightforward calculations,
we obtain the quadratic action for the tensor perturbations as
\begin{align}
  S^{T} = \frac{1}{2} \int \D \tau \frac{\D^3 k}{(2\pi)^3}
  \sum_{\pol=\pm}
  \left[\Delta_I^{\pol\prime\dagger} \Delta_{I}^{\pol\prime}
  + \Delta_I^{\pol\prime\dagger} K_{t,I J} \Delta_J^{\pol}
  - \Delta_I^{\pol\dagger} K_{t,I J} \Delta_J^{\pol\prime}
  - \Delta_I^{\pol\dagger} \Omega_{t,I J}^2 \Delta_J^{\pol}\right],
\end{align}
where $\tau$ is the conformal time defined by $\D\tau=\D t/a (t)$,
a dash denotes differentiation with respect to $\tau$, and 
\begin{align}
  \Delta^{\pol} = \mqty(h^{\pol} \\ T^{\pol}),
\end{align}
i.e. $\Delta^\sigma_1=h^\sigma$ and $\Delta^\sigma_2=T^\sigma$.
Here, $K_{t, IJ}$ is an anti-Hermitian matrix, and $\Omega^2_{t, IJ}$ is a Hermitian matrix, the explicit forms of which are found in Appendix~\ref{App: Perturbation}.

Following the quantization procedure given in Ref.~\cite{Nilles:2001fg} and Appendix~\ref{App: Quantaization},
we quantize $\Delta_{I}^{\pol}$
and express the operator $\hat{\Delta}_{I}^{\pol}$
in terms of the creation and annihilation operators
as
\begin{align}
  \hat{\Delta}_{I}^{\pol}(\tau, \bm{k})
  = \cald_{IJ}^{\pol}(\tau,k) \hat{a}_{J}^{\pol}(\bm{k})
  + \cald^{\pol *}_{IJ}(\tau,k) \hat{a}^{\pol\dagger}_{J}(\bm{k}),
\end{align}
where the commutation relations
\begin{align}
  [\hat{a}_{I}^{\pol}(\bm{k}),\hat{a}^{\pol'\dagger}_{J}(\bm{q})]
  = \delta^{\pol\pol'} \delta_{IJ} \delta (\bm{k}-\bm{q}),
  \quad \text{others} = 0,
\end{align}
are understood.
The equation of motion for the mode functions $\cald_{IJ}^{\pol}(\tau,k)$ are given by
\begin{align}
  \cald^{\pol\prime\prime}_{IL} + 2K_{t,IJ} \cald^{\pol\prime}_{JL}
  + \qty(K'_{t,IJ} + \Omega^2_{t,IJ} ) \cald_{JL}^{\pol} = 0.
  \label{eq: EOMforTensor}
\end{align}
We solve this equation numerically with the adiabatic vacuum initial conditions,
\begin{align}
  \cald^{\pol}_{IJ} = \frac{1}{\sqrt{2k}} c_{t,IJ}^{-1/2},
  \quad
  \cald^{\pol\prime}_{IJ} = -i \sqrt{\frac{k}{2}} c_{t,IJ}^{1/2},
\end{align}
where $c^2_{t, IJ}$ is a diagonal matrix element corresponding to the propagation speeds of the tensor modes.
The general form of the power spectrum is given by Eq.~\eqref{eq: PowerSpectrum} in Appendix~\ref{App: Quantaization}.
In the case of \acp{PGW}, we have
\begin{align}
  \calp_{h} (k) = \calp_{h}^{+} (k) + \calp_{h}^{-}(k),
  \quad
  \calp_{h}^{\pol} (k) = \frac{k^3}{2\pi^2}
  \frac{4}{a^2 \calg_{T}}
  \qty(|\cald_{11}^{\pol}|^2 + |\cald_{12}^{\pol}|^2 ),
  \label{eq: PowerGW}
\end{align}
where $\calg_{T}$ is defined in terms of the Horndeski functions as
\begin{align}
    \calg_{T}:=\mpl^2+2M^2\qty(
        F_4-2\hat XF_4'-\frac{H}{M}\frac{\dot\phi}{M^2}\hat XF_5'
  ).
\end{align}

\subsection{Scalar perturbations}

We move to derive the scalar power spectrum in the Horndeski-SU(2) system.
The basic procedure is essentially the same as that for the tensor modes.
After integrating out the nondynamical variables,
the quadratic action for the scalar perturbations takes the form of
\begin{align}
  S^{S} = \frac{1}{2} \int \D \tau \frac{\D^3 k}{(2\pi)^3}
  \left[\Delta_I^{\prime\dagger} \Delta_{I}^{\prime}
  + \Delta_I^{\prime\dagger} K_{s,I J} \Delta_J^{}
  - \Delta_I^{\dagger} K_{s,I J} \Delta_J^{\prime}
  - \Delta_I^{\dagger} \Omega_{s,I J}^2 \Delta_J^{}\right],
\end{align}
where
\begin{align}
  \Delta = \mqty(\dphi \\ \dQ \\ Z),
\end{align}
i.e. $\Delta_1=\dphi$, $\Delta_2=\dQ$, and $\Delta_3=Z$.
Here, $K_{s, IJ}$ is an anti-Hermitian matrix, and $\Omega^2_{s, IJ}$ is a Hermitian matrix, the explicit forms of which are found in Appendix~\ref{App: Perturbation}.

Following the same quantization procedure as in the case of the tensor perturbations,
we quantize $\Delta_{I}$ and write
\begin{align}
  \hat{\Delta}_{I}(\tau, \bm{k})
  = \cald_{IJ}(\tau,k) \hat{a}_{J}(\bm{k})
  + \cald^{*}_{IJ}(\tau,k) \hat{a}^{\dagger}_{J}(\bm{k}),
\end{align}
where the creation and annihilation operators satisfy
\begin{align}
  [\hat{a}_{I}(\bm{k}),\hat{a}^{'\dagger}_{J}(\bm{q})]
  = \delta_{IJ} \delta (\bm{k}-\bm{q}),
  \quad \text{others}=0.
\end{align}
We then numerically solve the equations of motion for the mode functions,
\begin{align}
  \cald^{\prime\prime}_{IL} + 2K_{t,IJ} \cald^{\prime}_{JL}
  + \qty(K'_{s,IJ} + \Omega^2_{s,IJ} ) \cald_{JL} = 0,
  \label{eq: EOMforScalar}
\end{align}
with the adiabatic vacuum initial conditions,
\begin{align}
  \cald_{IJ} = \frac{1}{\sqrt{2k}} c_{s,IJ}^{-1/2},
  \quad
  \cald^{\prime}_{IJ} = -i \sqrt{\frac{k}{2}} c_{s,IJ}^{1/2},
\end{align}
where $c^2_{s, IJ}$ is a diagonal matrix element corresponding to the propagation speed of each degree of freedom.
The general form of the power spectrum~\eqref{eq: PowerSpectrum} is now specialized to the present case, giving
\begin{align}
  \calp_{\dphi} (k) = \frac{k^3}{2\pi^2}
  \frac{4}{a^2 \cala_{S}}
  \qty(|\cald_{11}|^2 + |\cald_{12}|^2 + |\cald_{13}|^2),
  \label{eq: PowerAxion}
\end{align}
where
\begin{align}
  \cala_{S}&:=
  1+F_2'+2\hat XF_2''+6\frac{H}{M}\frac{\dot\phi}{M^2}\left(F_3'+\hat XF_3''\right)
  +6\frac{H^2}{M^2}\left(F_4'+8\hat X F_4''+4\hat X^2F_4'''\right)
  \notag \\ &\quad 
  +\frac{H^3}{M^3}\frac{\dot\phi}{M^2}
  \left(6F_5'+14\hat XF_5''+4\hat X^2F_5'''\right).
\end{align}

We evaluate the power spectrum of the curvature perturbation on uniform density slices $\zeta$.
The axion-field perturbation, $\delta\phi$, gives the dominant contribution to $\zeta$,
leading to~\cite{Dimastrogiovanni:2012ew, Dimastrogiovanni:2023oid}
\begin{align}
  \zeta \simeq -\frac{H}{\dot{\phi}} \dphi.
  \label{eq: defCurvaturePtb}
\end{align}
(This is true in the present setup as well as in the previous models, as discussed in more detail in Appendix~\ref{App: CurvaturePtb}.)
The power spectrum of $\zeta$ is thus given by 
\begin{align}
  \calp_{\zeta}
  = \frac{k^3}{2 \pi^2} \frac{H^2}{\dot{\phi}^2}
  \calp_{\dphi}
  .
  \label{eq: PowerZeta}
\end{align}

We have thus derived the basic equations for evaluating the tensor and scalar power spectra in the Horndeski-SU(2) system.
Using those results, we can now compute the gravitational wave spectra and the CMB observables for the enhanced friction models of CNI.

\section{Gravitational wave spectra and CMB observables} \label{sec: Observation}

In this section, we discuss how the choice of the Horndeski functions in the enhanced friction models of CNI affects the gravitational wave power spectra and the CMB observables and look for the models that can generate observable gravitational waves on small scales while avoiding their overproduction in the CMB scale.

\subsection{Power spectra}

\begin{figure}
  \begin{center}
    \includegraphics[keepaspectratio=true,width=0.49\textwidth]{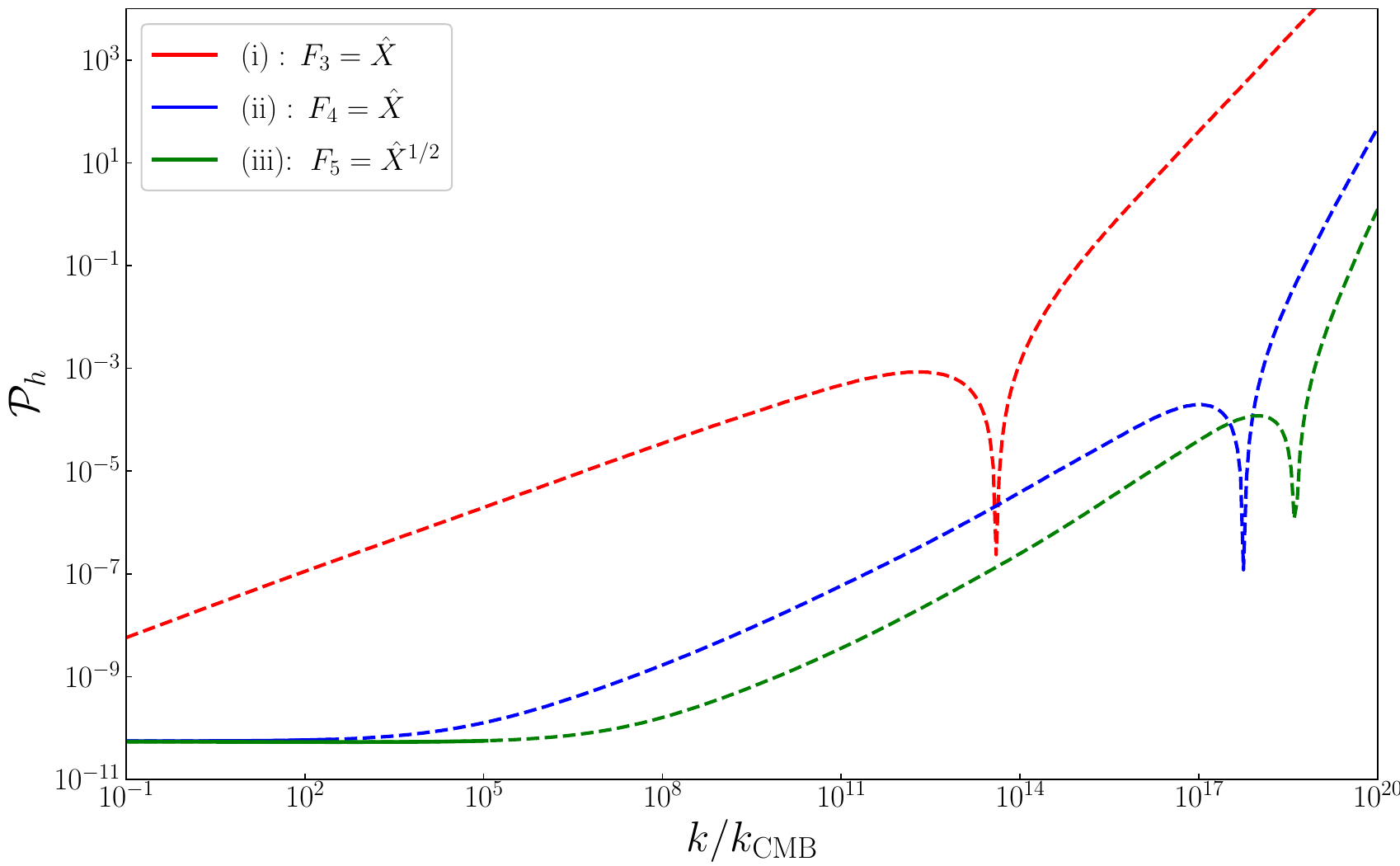}
    \includegraphics[keepaspectratio=true,width=0.49\textwidth]{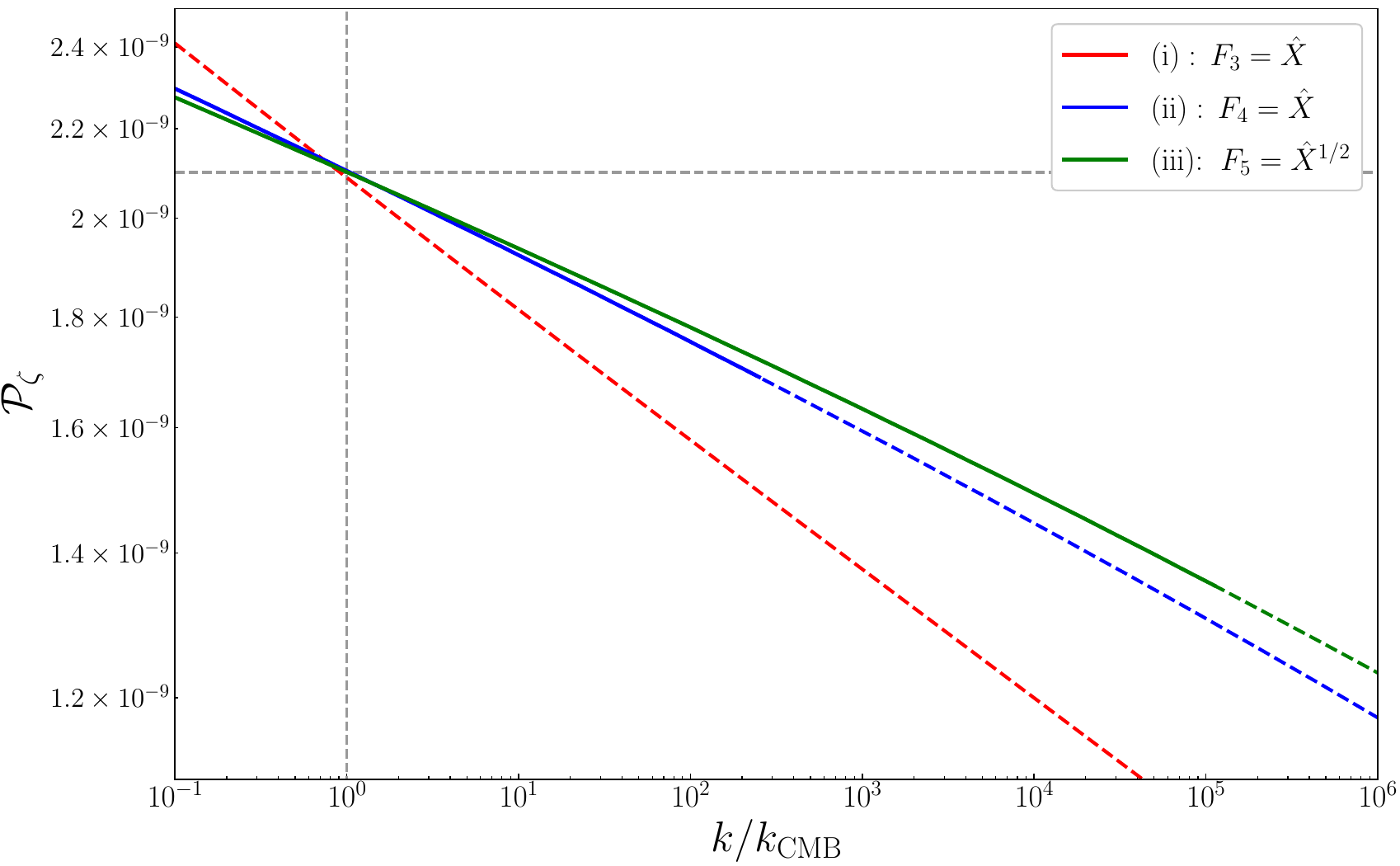}
  \end{center}
  \caption{Power spectra of the \ac{PGW} (left) and
  the curvature perturbation (right) as functions of $k/k_{\rm CMB}$.
  Parameters are given by Eq.~\eqref{eq: FiducialParams}.
  Different lines correspond to different choices of the Horndeski functions.
  In the regime denoted by solid lines, backreaction effects are negligible.
  However, significant backreaction is anticipated in the regime indicated by dashed lines.
  In the right panel, the vertical dashed line shows the CMB scale ($k=k_{\textrm{CMB}}$)
  and the horizontal dashed line stands for $\calp_{\zeta}=2.1\times 10^{-9}$.
  }
  \label{fig: plot_PS}
\end{figure}

We show in the left panel of Fig.~\ref{fig: plot_PS} the power spectra of \acp{PGW} for the three different models introduced in Sec.~\ref{subsec:models}.
The parameters are the same as those used in Fig.~\ref{fig: DensityParamBG}.
It can be seen that the power spectrum is enhanced for wavenumbers larger than some $k=k_\star$ due to the gauge field.
We find that the mode with $k_\star$ exits the horizon at the time when $\Omega_B$ exceeds $\Omega_X$.
The precise value of $k_\star$ thus depends on the choice of the Horndeski functions, as can be seen from Fig.~\ref{fig: DensityParamBG}.
In the model (i), we see that $\Omega_B>\Omega_X$ already at 70 e-folds before the end of inflation, resulting in the overproduction of \ac{PGW}, similarly to the conventional \ac{CNI} model.
The tensor power spectrum is, therefore, too large to be consistent with observations at the \ac{CMB} scale.
In contrast, we have $\Omega_B\sim\Omega_X$ in the model (ii) and $\Omega_B<\Omega_X$ in the model (iii) at 70 e-folds before the end of inflation.
In these two cases, the tensor power spectrum is evaluated to be $\calp_{h} \sim 7\times 10^{-11}$, leading to the tensor-to-scalar ratio $r \sim 0.03$, which is consistent with the \ac{CMB} observations at $2\sigma$~\cite{BICEP:2021xfz}.
(A more detailed comparison with observations, including the scalar spectral index, will be performed shortly.)
One can understand these results by recalling how the value of the gauge field trapped at the minimum of the effective potential, $Q_{\rm min}$, depends on $\dot\phi$ [Eq.~\eqref{eq: QandPhi}].
The more effective enhanced friction is, the smaller $\dot\phi$ is and so is $Q_{\rm min}$. The growth of the gauge field is thus suppressed by enhanced friction.

In the right panel of Fig.~\ref{fig: plot_PS}, we show the power spectra of the curvature perturbation $\zeta$.
In the model (i), the tilt of the spectrum is negatively large compared to the other two models. 
This result reminds us of the original \ac{CNI} model, where it has been known that $n_s$ is negatively large~\cite{Adshead:2013nka}.
It implies that the effects of enhanced friction on the scalar perturbations are almost negligible in the model (i), yielding results close to those of the original \ac{CNI} model.

\begin{figure}
  \begin{center}
    \includegraphics[keepaspectratio=true,width=0.49\textwidth]{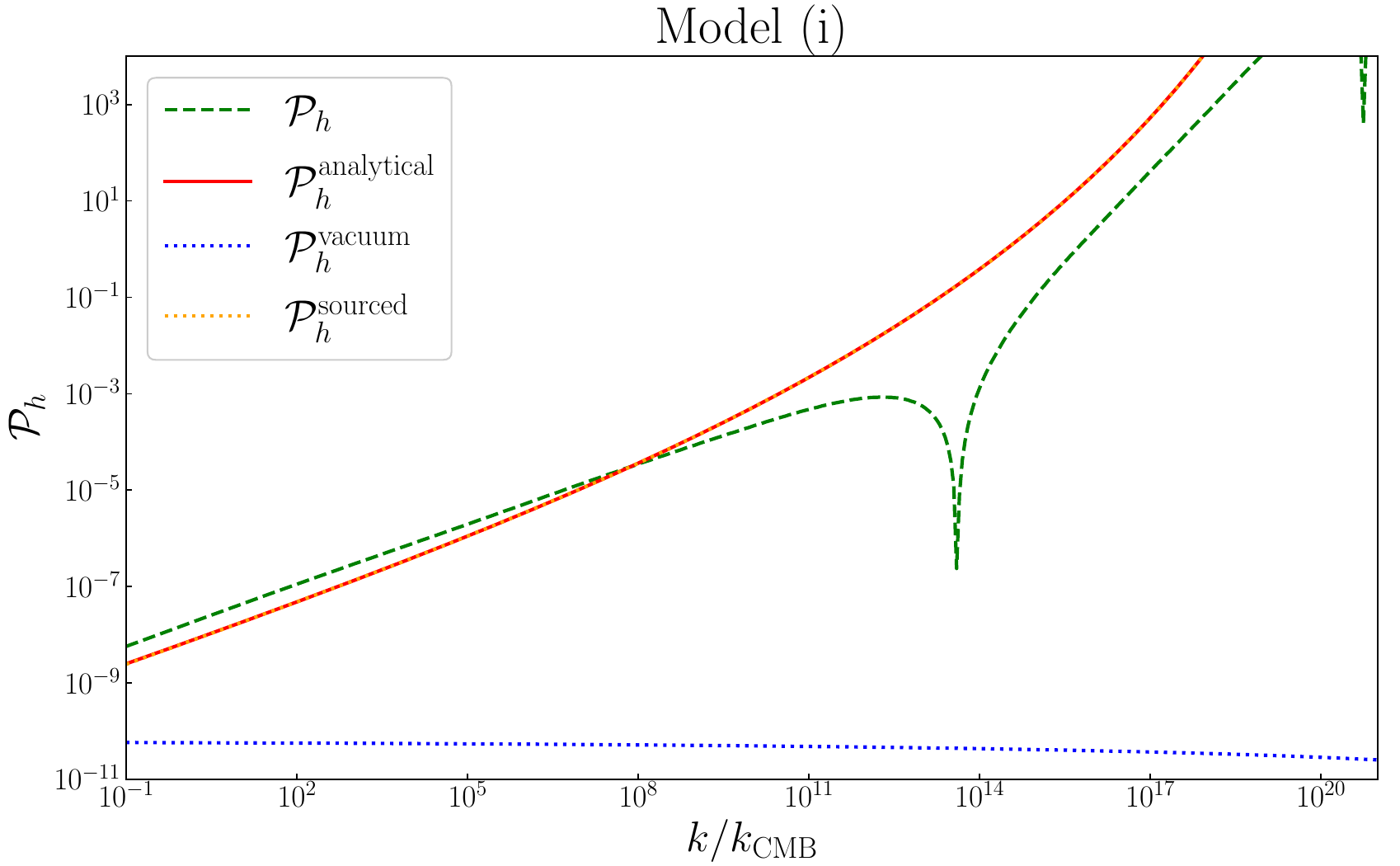}
    \includegraphics[keepaspectratio=true,width=0.49\textwidth]{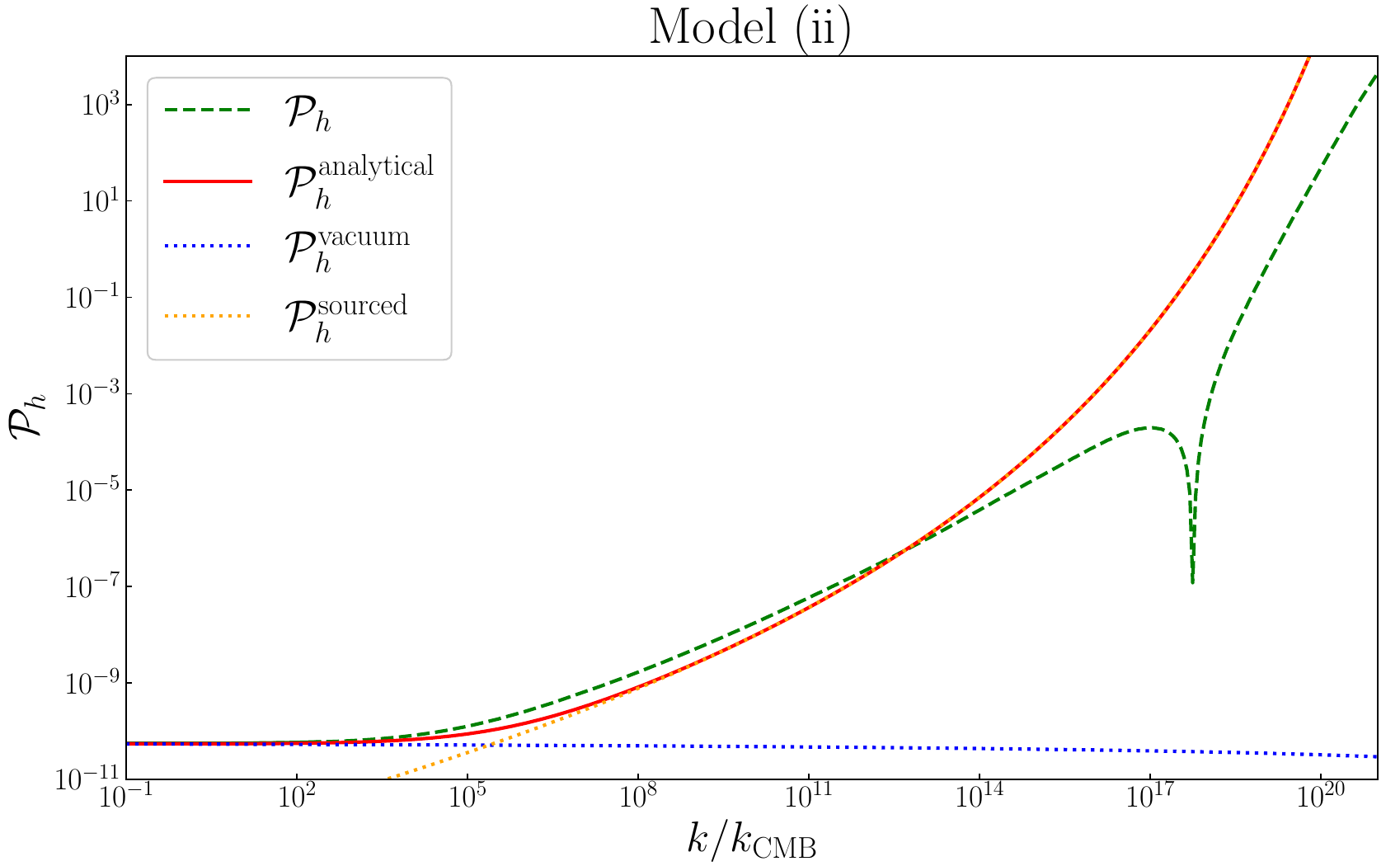}
    \includegraphics[keepaspectratio=true,width=0.49\textwidth]{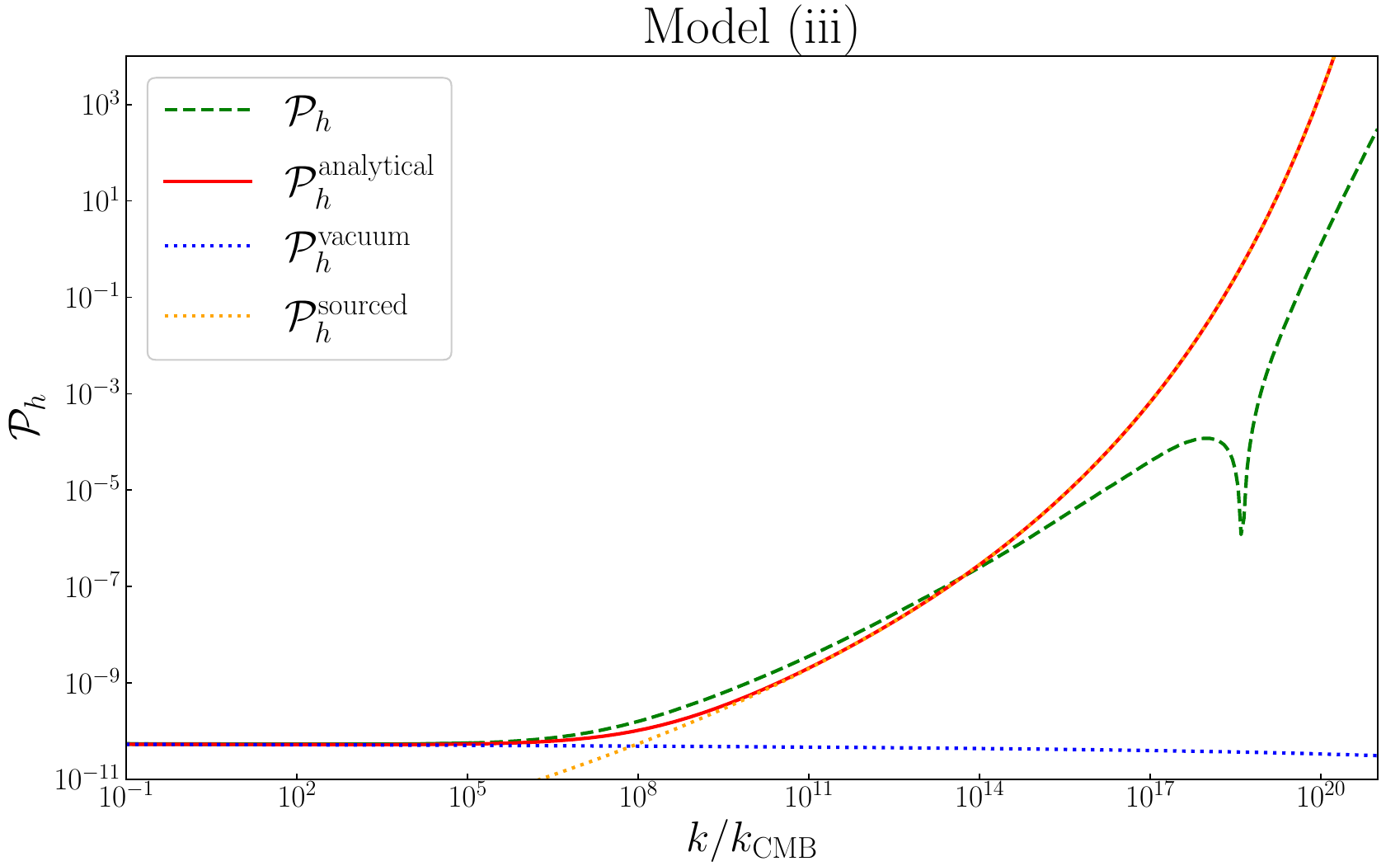}
  \end{center}
  \caption{
  Analytically and numerically obtained power spectra of \ac{PGW} as functions of $k/k_{\rm CMB}$.
  Parameters are given by Eq.~\eqref{eq: FiducialParams}.
  Each panel corresponds to different choices of the Horndeski functions.
  Green dashed lines show the numerical results.
  Blue (orange) dotted lines represent the analytic results obtained from Eq.~\eqref{eq: calP_v} (Eq.~\eqref{eq: calP_s}),
  and red solid lines show the sum of the two expressions~\eqref{eq: calP_all}.
  }
  \label{fig: plot_PSanalytic}
\end{figure}

Let us derive the analytic expression for the power spectrum of \acp{GW}.
In the early stage of inflation, the effects of enhanced friction are dominant,
while those of the gauge field can be ignored.
The power spectrum is therefore well approximated by the expression for slow-roll inflation in Horndeski gravity derived in~\cite{Kobayashi:2011nu}.
In the present case, we have
\begin{align}
  \calp_{h}^{\rm vacuum}
  =8
  \frac{\left[\mpl^2+2M^2\left(F_4-2\hat XF_{4\hat X}-
  \frac{H}{M}\frac{\dot\phi}{M^2}\hat XF_{5\hat X}\right)\right]^{1/2}}%
  {\left[\mpl^2+2M^2\left(F_4-\frac{\ddot\phi}{M^3}\hat X F_{5\hat X}\right)\right]^{3/2}}
  \left.\qty(\frac{H}{2\pi})^2\right|_{k=aH}.
  \label{eq: calP_v}
\end{align}
This expression is mainly relevant to large scales with $k<k_\star$.
In the late stage, the gauge field is active, while the effects of enhanced friction can be ignored.
Therefore, the power spectrum can be approximated by the formula obtained in the original \ac{CNI} model~\cite{Dimastrogiovanni_2017}:
\begin{align}
  \calp_{h}^{\rm sourced} = \left.\frac{\epsilon_{B} H^2}{\pi^2 \mpl^2}
  e^{3.6 m_{Q}}\right|_{k=aH}.
  \label{eq: calP_s}
\end{align}
where $\epsilon_{B}=\ga^2 Q^4/(\mpl^2 H^2)$ and $m_{Q}=\ga Q/H$.
This expression is mainly relevant to small scales with $k>k_\star$.
Thus, our analytic expression for the \ac{GW} spectrum is given by
\begin{align}
  \calp_{h}^{\rm analytic} = \calp_{h}^{\rm vacuum} + \calp_{h}^{\rm sourced},
  \label{eq: calP_all}
\end{align}
where the first (second) term is dominant for $k<k_\star$ ($k>k_\star$).
In Fig.~\ref{fig: plot_PSanalytic}, we compare the numerical results with the analytic expression, which justifies the above argument.

Now we have a simplified way of deriving the approximate \ac{GW} spectrum that can be used for any enhanced friction model of \ac{CNI}.
First, one solves the background cosmological evolution to identify the time when $\Omega_B=\Omega_X$ occurs, which fixes the wavenumber $k_\star$.
Then, the approximate \ac{GW} spectrum is given by Eq.~\eqref{eq: calP_v} for $k<k_*$ and Eq.~\eqref{eq: calP_s} for $k>k_\star$.
The tensor spectral index on small scales is given by
\begin{align}
    n_t\simeq 3.6\left(\epsilon+\frac{\dot Q_{\rm min}}{HQ_{\rm min}}\right)m_Q.
\end{align}

\subsection{CMB observables}

Let us now calculate the scalar spectral index and the tensor-to-scalar ratio,
\begin{align}
  \ns = 1 + \dv{\ln{\calp_{\zeta}}}{\ln{k}},
  \quad
  r = \frac{\calp_{h}}{\calp_{\zeta}},
\end{align}
for different choices of the Horndeski functions and different parameters.
We evaluate $\ns$ in the \ac{CMB} scale $k=k_{\rm CMB}$ that exits the horizon 60 e-folds before the end of inflation.
We use $r_{0.002}:=r(k_{\rm CMB}/25)$, which corresponds to the pivot scale $0.002\,{\rm Mpc}^{-1}$.

\begin{figure}
  \begin{center}
    \includegraphics[keepaspectratio=true,height=80mm]{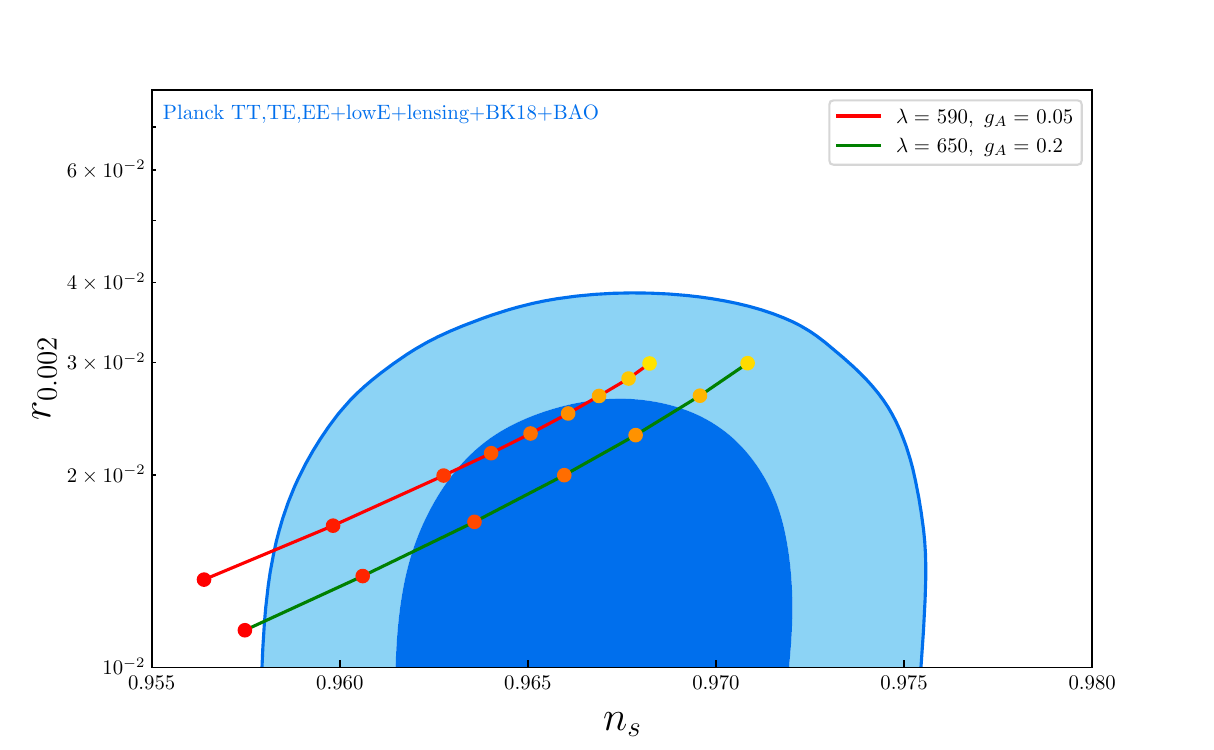}
  \end{center}
  \caption{Points in the $n_s$-$r$ plane in the case of the model (iii).
  The $1\sigma$ and $2\sigma$ contours are taken from the Planck+BICEP/Keck constraints~\cite{BICEP:2021xfz}.
  The color shows the value of $\mu$.
  }
  \label{fig: BK18_nsr_G5X2}
\end{figure}

We first consider the model (iii) ($F_{5}=\hat{X}^{1/2}$) because this model can avoid the overproduction of gravitational waves at the \ac{CMB} scale as discussed above
(at least for the parameters given in Eqs.~\eqref{eq: FiducialParams} and~\eqref{eq:paras2}).
In Fig.~\ref{fig: BK18_nsr_G5X2}, we show $n_s$ and $r$ for different sets of the parameters.
More specifically, for $(f,\ga, \lambda)=(0.2, 0.05, 590)$ and $(f,\ga, \lambda)=(0.2, 0.2, 650)$,
we vary $\mu$ in the range $4.2\times 10^{-3}\leq\mu\leq 5.0\times 10^{-3}$
and thereby explore the viable parameter space.
The value of $M$ is fixed accordingly by demanding that $\calp_{\zeta}=2.1\times 10^{-9}$.
It can be seen from Fig.~\ref{fig: BK18_nsr_G5X2} that almost the entire parameter space we have investigated is consistent with Planck$+$BICEP/Keck constraints at the $1\sigma$ level.
This should be contrasted with the results of the model (ii) studied in Ref.~\cite{Dimastrogiovanni:2023oid}:
the model (ii) is consistent with Planck$+$BICEP/Keck constraints at the $2\sigma$ level.

\begin{figure}
  \begin{center}
    \includegraphics[keepaspectratio=true,height=80mm]{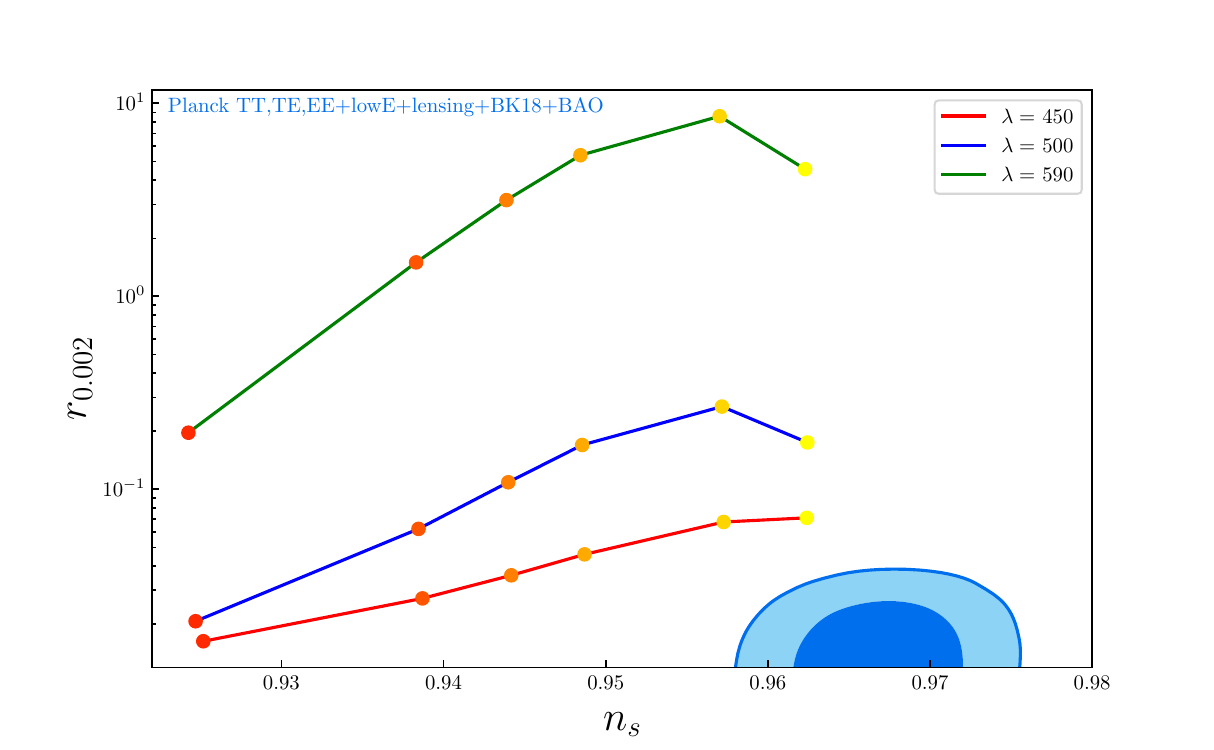}
  \end{center}
  \caption{Points in the $n_s$-$r$ plane in the case of the model (i).
  The $1\sigma$ and $2\sigma$ contours are taken from the Planck+BICEP/Keck constraints~\cite{BICEP:2021xfz}.
  The color of the circle shows the value of $\mu$.
  }
  \label{fig: BK18_nsr_G3X}
\end{figure}

We then study the model (i) ($F_3=\hat X$).
In Fig.~\ref{fig: BK18_nsr_G3X}, we show our results for
$f=0.2\mpl$, $\ga=0.05$, $\lambda=450,500,590$, and $4.2\times 10^{-3}\leq\mu\leq 6\times 10^{-3}$.
It is found that in the entire parameter space we investigated, $n_s$ is too small and/or $r$ is too large to be consistent with observations.

\subsection{Gravitational waves and backreaction}

\begin{figure}
  \begin{center}
    \includegraphics[keepaspectratio=true,height=80mm]{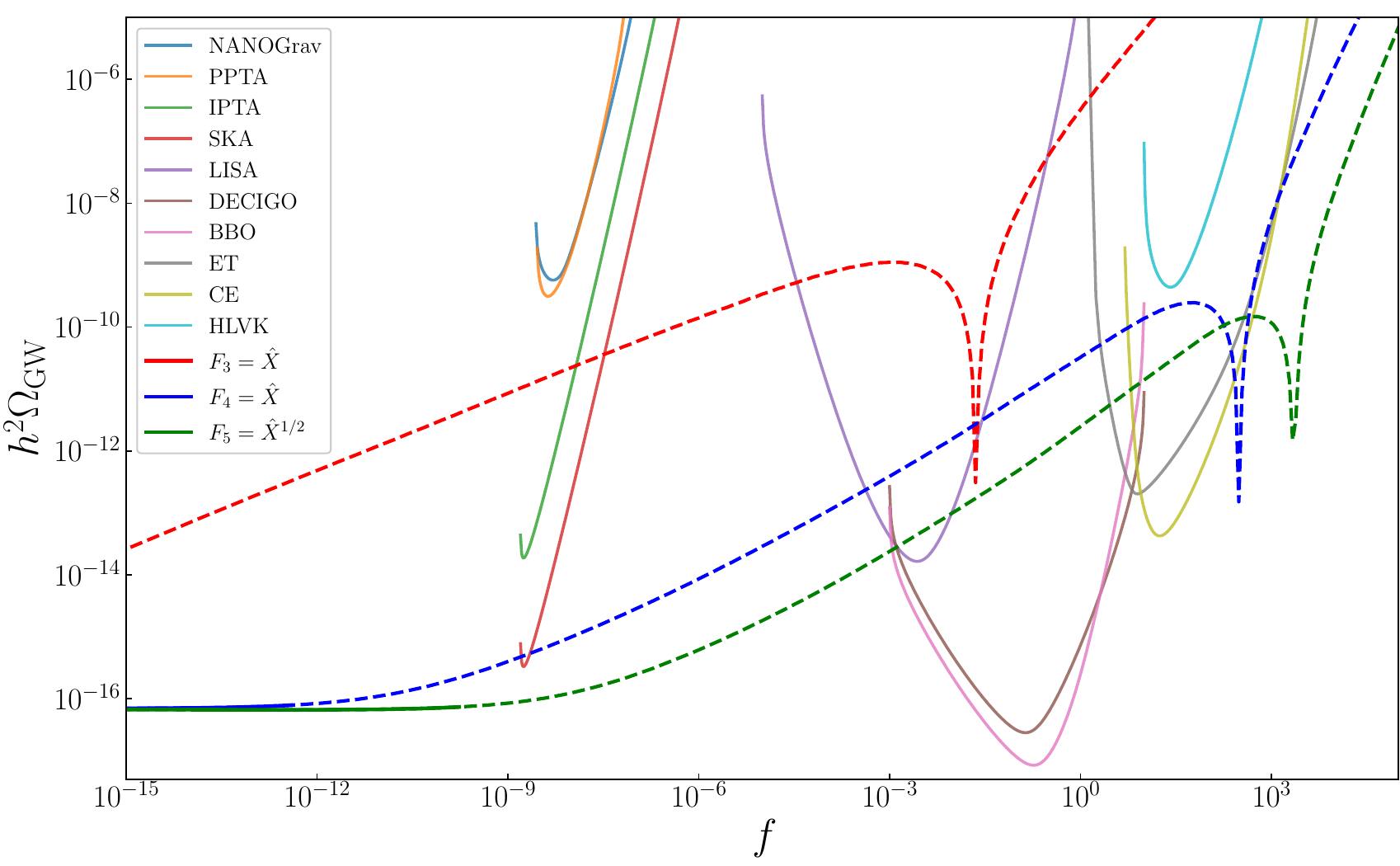}
  \end{center}
  \caption{Plot of $\Omega_{\rm GW}$.
  Solid lines show results with an insignificant backreaction regime.
  Dashed lines show the backreaction regime that satisfies Eq.~\eqref{eq: BRconstraint}.
  The sensitivity curve is obtained from Ref.~\cite{Schmitz:2020syl}.
  }
  \label{fig: OmegaGW}
\end{figure}

Let us move to discuss the detectability of \acp{PGW} from the enhanced friction models of CNI.
The density parameter of \acp{GW} per log frequency interval is given by
\begin{align}
  \Omega_{\rm GW} (k) = \frac{3}{128} \Omega_{r,0} \calp_{h}
  \qty[\frac{1}{2} \qty(\frac{k}{k_{\rm eq}})^2 + \frac{16}{9} ],
\end{align}
where $\Omega_{r,0}\simeq 2.47 \times 10^{-5}$ is the present radiation density parameter and $k_{\rm eq}\simeq 0.013 \,{\rm Mpc}^{-1}$ is the wavenumber of the mode reentering the horizon at matter-radiation equality with $\Omega_{m}h^2\simeq 0.141$ (see e.g.~\cite{Caprini:2018mtu}).
We express $\Omega_{\rm GW}$ as a function of frequency by using the relation $f \simeq 1.5 \times 10^{-15}(k/\rm{Mpc^{-1}})\,$Hz.
In Fig.~\ref{fig: OmegaGW}, we present $\Omega_{\rm GW}$ for the models (i), (ii), and (iii).
The parameters are given again by Eqs.~\eqref{eq: FiducialParams} and \eqref{eq:paras2}.
The sensitivity curves are taken from Ref.~\cite{Schmitz:2020syl}.
As we already discussed, the amplitude of GWs in the model (i) is too large on the CMB scale.
Therefore, our main interest here is in the models (ii) and (iii).
Figure~\ref{fig: OmegaGW} shows that \acp{PGW} in these models would be detectable by 
experiments such as SKA, LISA, and DECIGO/BBO.
The SKA might be capable of observing circular polarization~\cite{Cruz:2024esk}.
If the circular polarization of the \acp{PGW} were detected, our result could provide the physical interpretation of the observational results.

Some comments are now in order.
The first one is on the backreaction from tensor perturbations.
In the axion-SU(2) inflation model, tensor perturbations of the SU(2) gauge field are amplified by the tachyonic instability slightly before the horizon crossing.
These large tensor perturbations cause the backreaction to the background dynamics (see Refs.~\cite{Fujita:2017jwq, Ishiwata:2021yne, Iarygina:2023mtj, Dimastrogiovanni:2024xvc} for detailed discussions).
The effects of the backreaction are significant when $Q$ becomes large enough to satisfy~\cite{Fujita:2017jwq}
\begin{align}
  e^{1.85 m_{Q}} \ga \gtrsim 10,
  \label{eq: BRconstraint}
\end{align}
where $m_{Q}=\ga Q/H$.
This condition can also be applied to the present model at late times when gravitational waves are sourced by tensor perturbations of the gauge field.
This backreaction may change the scalar and tensor spectra on small scales.
In Figs.~\ref{fig: plot_PS} and~\ref{fig: OmegaGW}, we show by dashed lines the regimes where Eq.~\eqref{eq: BRconstraint} is satisfied.

The backreaction effects are likely insignificant on large scales.
Therefore, the predictions on CMB observables, $n_s$ and $r$, are robust even when the effects of the backreaction are taken into account.
The issue needs to be revisited when a more detailed treatment of backreaction becomes feasible in the future.
Further research such as lattice simulations is anticipated to advance our understanding of the impact of backreaction.

The second comment is on the perturbativity constraints on the axion-gauge dynamics.
Very recently, the authors of Ref.~\cite{Dimastrogiovanni:2024lzj} put constraints on the allowed parameter space of axion-gauge inflation by requiring that loop corrections to the gauge-field propagator remain small.
The resultant constraints are competitive with the backreaction bounds and relevant to the present models as well.

Finally, the third comment is on $\ga$.
The parameter $\ga$ is crucial in axion-gauge field interactions during inflation. 
The dissipation term of the axion (the right-hand side of Eq.~\eqref{eq: EOMphi}) is proportional to $\ga^{-2}$, as we see from Eq.~\eqref{eq: QandPhi} the relation $Q \propto \ga^{-1}$ (for nearly constant $\dot\phi$ and $H$). Therefore, for smaller $\ga$, the dissipation from the axion to the gauge field is more efficient.
As a result,
for sufficiently small $\ga$, enhanced friction is less effective than the interaction with the gauge field in the axion field equation and hence it does not effectively suppress the growth of the gauge field. 
For instance, for $\ga=10^{-3}$ (used in Ref.~\cite{Fujita:2022jkc}), the effects of enhanced friction are negligible, resulting in a large tensor-to-scalar ratio at the \ac{CMB} scale.
On the other hand, if $\ga$ is large enough, the aforementioned backreaction problem could arise. 
Thus, a careful choice of the value of $\ga$ is important for building workable models of the axion-gauge field interaction during inflation and making reliable predictions that can be tested by future observations.

\section{Conclusions} \label{sec: conclusion}

In this paper, we have generalized chromo-natural inflation (CNI) to allow the axion to be coupled nonminimally to gravity through the Horndeski-type interaction, extending the previous model~\cite{Dimastrogiovanni:2023oid} to a more general, unifying framework.
Due to the overproduction of gravitational waves (GWs), the original (minimally coupled) \ac{CNI} model has already been ruled out by \ac{CMB} observations~\cite{Adshead:2013nka}.
Nonminimal couplings to gravity can help to avoid this inconsistency with \ac{CMB} observations by reducing the production of \acp{GW} on large scales.
In the late stage of inflation, the effects of the SU(2) gauge field overwhelm this reduction mechanism and enhance the \ac{GW} spectrum on small scales.

Inspired by~\cite{Dimastrogiovanni:2023oid}, we have considered general shift-symmetric Horndeski-type interactions of the axion field $\phi$ and studied how they contribute to the enhancement factor $\cali$ of the friction term in $\phi$'s equation of motion.
The enhancement factor typically has the form
$\cali\sim H^p\dot\phi^q$, and thus is characterized by the powers $p$ and $q$.
We have studied the background slow-roll dynamics of Horndeski-SU(2) inflation for several representative models with different $(p,q)$ including the one proposed in Ref.~\cite{Dimastrogiovanni:2023oid}.
We have also considered cosmological perturbations and obtained the power spectra of \acp{GW} and the curvature perturbation for each model.
Through these analyses, we have found that the two density parameters $\Omega_{B}$ and $\Omega_{X}$ defined in Eq.~\eqref{eq: DensityParam} are of particular importance in determining the shape of the \ac{GW} spectrum.
In the early stage of inflation, one has $\Omega_{B}<\Omega_{X}$ and enhanced friction is effective.
The \ac{GW} production is suppressed in this stage.
The gauge field grows toward the end of inflation,
and hence one has $\Omega_{B}>\Omega_{X}$ in the late stage of inflation.
Now enhanced friction is less effective and the inflationary dynamics becomes similar to that of usual \ac{CNI}, leading to the efficient production of \acp{GW} on small scales.
This is a universal nature of nonminimally coupled \ac{CNI} models irrespective of $(p,q)$.
However, the time when $\Omega_B$ catches up with $\Omega_X$ crucially depends on $(p,q)$.
We have found that the onset of the late stage with $\Omega_B>\Omega_X$ is delayed for larger $p$ and smaller $q$.
For the \ac{GW} spectrum on large scales to be consistent with \ac{CMB} observations, the friction-dominated stage must be sustained sufficiently long.
Therefore, the model with $(p,q)=(3,0)$ is most appropriate.

We have further investigated the observational predictions of each model in more detail by evaluating the scalar spectral index $n_s$ and the tensor-to-scalar ratio $r$ and looked for regions of the parameter space consistent with \ac{CMB} observations.
We have found that the model with $(p,q)=(3,0)$ (the model (iii)) is consistent with Planck$+$BICEP/Keck constraints at the $1\sigma$ level, which is better than the results of the previous model~\cite{Dimastrogiovanni:2023oid}.
While consistent with observations on large scales, 
the tensor amplitude is enhanced on small scales, which could be relevant to future \ac{GW} experiments.
Though this could be an interesting observational signature, one must be careful about the issue of possible strong backreaction to the background dynamics caused on the same small scales.
Note that the predictions about $n_s$ and $r$ are not affected by this backreaction issue.

It would be interesting to study the reheating stage after inflation in the preset setup.
The nonminimal coupling considered in this paper leads to distinctive dynamics of the scalar field during reheating~\cite{Sato:2020ghj}, which would be helpful to distinguish between different enhanced friction models of inflation.
We hope to come back to this point in a future publication.

\acknowledgments
We thank Martino Michelotti for fruitful discussions.
The work of T.M. was supported by JSPS KAKENHI Grants No.~JP23KJ2007.
The work of T.K. was supported by
JSPS KAKENHI Grant No.~JP20K03936 and
MEXT-JSPS Grant-in-Aid for Transformative Research Areas (A) ``Extreme Universe'',
No.~JP21H05182 and No.~JP21H05189.

\appendix

\section{Background equations} \label{App: Background}
This appendix summarizes the background field equations for the cosmology of the Horndeski-SU(2) system described by the action~\eqref{action: HorndeskiSU(2)}.
In order to avoid unnecessarily messy expressions, we assume that $G_3$, $G_4$, and $G_5$ depend only on $X$, as is the case of the inflation models we study.

Substituting the homogeneous metric~\eqref{eq: FLRWmetric}, the scalar field $\phi=\phi(t)$, and the ansatz for the gauge field~\eqref{eq: GFansatz} to the action~\eqref{action: HorndeskiSU(2)},
we obtain the minisuperspace action in terms of $N(t)$, $a(t)$, $\phi(t)$, and $Q(t)$.
After deriving the field equations, we set $N=1$.
Variation with respect to $N(t)$ yields the Friedmann equation,
\begin{align}
    {\cal E} = - 3\mpl^2H^2 + \rho_\phi+\rho_Q,
\end{align}
with 
\begin{align}
    \rho_\phi&:=- G_{2} + 2 X G_{2X}
  + 6 H \dot{\phi} X G_{3X}
  +3\left(\mpl^2-2G_4\right)H^2
  + 24 H^2 X \qty(G_{4X} + X G_{4XX} )
  + 2 H^3 \dot{\phi} X \qty(5 G_{5X} + 2 X G_{5XX} ),
    \\ 
    \rho_Q&:=\frac{3}{2}\qty(\dot{Q} + HQ )^2
  + \frac{3}{2} \ga^2 Q^4.
\end{align}
Here we used the notation $G_{2X}=\partial G_2/\partial X$.
Variation with respect to $a(t)$ gives 
\begin{align}
    P = \mpl^2\left(3H^2+2\dot H\right) + p_\phi+p_Q,
\end{align}
where 
\begin{align}
    p_\phi&:=G_{2}
  + 3 H^2 \left(2G_{4}-\mpl^2\right)
  - 12 H^2 X G_{4X}
  - 4 H^3 \dot{\phi} X G_{5X}
  +\ddot\phi {\cal B}_{\phi}
  +2\dot H\left(\calg_T-\mpl^2\right),
    \\ 
    p_Q&:=\frac{1}{3}\rho_Q,\label{rad-eos}
\end{align}
with 
\begin{align}
  {\cal B}_{\phi} &:=
  - 2 X G_{3X}
  - 4 H \dot{\phi} G_{4X}
  - 8 H \dot{\phi} X G_{4XX}
  - 6 H^2 X G_{5X}
  - 4 H^2 X^2 G_{5XX},
  \\
  \calg_T &:=2\left(G_4-2XG_{4X}-HX\dot\phi G_{5X}\right)
  \label{eq: calgT}.
\end{align}
Note here that to obtain Eq.~\eqref{rad-eos} we used the field equation for $Q$,
which takes the same form as in conventional CNI,
\begin{align}
    \frac{{\cal M}_{Q}}{3} := \ddot{Q} + 3H\dot{Q} + (\dot{H} + 2H^2) Q + 2 \ga^2 Q^3
  - \frac{\ga \lambda}{f} \dot{\phi} Q^2=0.
\end{align}
Finally, the field equation for $\phi$ is given by 
\begin{align}
    {\cal M}_{\phi} := \frac{\D}{\D t}\left(\cali \dot\phi\right) 
    +3H\cali \dot\phi -G_{2\phi}+\frac{3g_A\lambda}{f}Q^2\left(\dot Q+HQ\right)=0,
\end{align}
where 
\begin{align}
    \cali&:=G_{2X}
  + 3 H \dot{\phi} G_{3X}
  + 6 H^2 G_{4X}
  + 12 H^2 X G_{4XX}
  + 3 H^3 \dot{\phi} G_{5X}
  + 2 H^3 \dot{\phi} X G_{5XX}.
\end{align}

\section{On the approximation of ignoring the scalar metric perturbations}\label{app: MetricPtb}

In the main text, we ignore scalar metric perturbations in the flat gauge to reduce the computational cost.
In this appendix, we discuss the validity of this approximation by comparing the results obtained by ignoring the scalar metric perturbations with those obtained with no such approximations.

The full expression for the quadratic action for the scalar modes is obtained as
\begin{align}
  S_{\mathrm{S}}^{(2)}=\int \dd[4]{x} a^3
  & \bigg[ \Sigma \dN^2
  - \frac{\partial \mathcal{E}}{\partial \dot{\phi}} \ddphi \dN
  - \frac{\partial \mathcal{E}}{\partial \dot{Q}}
  \qty(\ddQ + \frac{1}{3}\partial^2 \dot{Z} - \frac{1}{3} \partial^2 Y ) \dN
  - \frac{\partial \mathcal{E}}{\partial \phi} \dphi \dN
  - \frac{\partial \mathcal{E}}{\partial Q}
  \qty(\dQ + \frac{1}{3}\partial^2 Z ) \dN
  \notag\\
  &
  - \frac{{\cal B}_{\phi}}{a^2} \partial^2 \dphi \dN
  - 2 \Theta \dN \frac{\partial^2 B}{a^2}
  + \cali \dot{\phi} \dphi \frac{\partial^2 B}{a^2}
  + \mathcal{B}_{\phi} \ddphi \frac{\partial^2 B}{a^2}
  + \frac{2}{3} \frac{\partial {\cal E}}{\partial \dot{Q}} \dQ \frac{\partial^2 B}{a^2}
  + \frac{\ga^2 Q^4}{a^2} (\partial B)^2
  \notag\\
  &
  + 2 \ga^2 a^2 Q^3 Y \frac{\partial^2 B}{a^2}
  + \frac{1}{2} (\partial^2 Y)^2
  + \ga^2 a^2 Q^2 (\partial Y)^2
  - \qty( \ddQ + \partial^2 \dot{Z}
  + H \dQ + H \partial^2 Z
  - \frac{\ga \lambda Q^2}{f} \dphi ) \partial^2 Y
  \notag\\
  &
  + \frac{3}{2} (\ddQ)^2
  - \frac{1}{a^2} (\partial \dQ)^2
  + \frac{1}{2} \frac{\partial {\cal M}_{Q}}{\partial Q} \dQ^2
  + \frac{1}{2} (\partial^2 \dot{Z})^2
  - \frac{1}{2} \qty(2 \ga^2 Q^2 + \dot{H} + 2 H^2 ) (\partial^2 Z)^2
  \notag\\
  &
  + \ddQ \partial^2 \dot{Z}
  - \frac{\partial {\cal M}_{Q}}{\partial Q} \dQ \partial^2 Z
  + \frac{\ga \lambda Q^2}{f}
  \qty(\ddphi \partial^2 Z
  - 3 \dphi \ddQ )
  - \frac{3 \ga \lambda Q}{f}
  \qty(2 \dot{Q} + 3 H Q ) \dphi \dQ
  \notag\\
  &
  + \frac{1}{2} \cala_{S} \ddphi^2
  - \frac{1}{2 a^2} \calb_{S} \partial_i \dphi \partial^i \dphi
  - \frac{1}{2} {\cal C}_{S} \dphi^2
  \bigg],
\end{align}
where
\begin{align}
  &\Sigma = \frac{1}{2}
  \left(
    \dot{\phi} \frac{\partial {\cal E}}{\partial \dot{\phi}}
  + H \frac{\partial {\cal E}}{\partial H}
  + \dot{Q} \frac{\partial {\cal E}}{\partial \dot{Q}}
  \right),
  \quad
  \Theta = -\frac{1}{6} \frac{\partial {\cal E}}{\partial H}
  + \frac{1}{6} Q \frac{\partial {\cal E}}{\partial \dot{Q}}
  \quad
  {\cala}_{S}  =
  \cali
  + \dot{\phi}\frac{\partial \cali}{\partial \dot{\phi}},
  \quad
  {\cal C}_{S} = \frac{\partial {\cal M}_{\phi}}{\partial \phi},
\end{align}
and
\begin{align}
  \calb_{S} &=
  G_{2X}
  + 2 \ddot{\phi} \qty(G_{3X} + X G_{3XX} )
  + 4 H \dot{\phi} G_{3X}
  \notag\\
  &\quad
  + \qty(6 H^2 + 4 \dot{H} ) G_{4X}
  + \qty(20 H^2 X
  + 8 \dot{H} X
  + 12 H \dot{\phi} \ddot{\phi} ) G_{4XX}
  + 8 H \dot{\phi} \ddot{\phi} X G_{4XXX}
  \notag\\
  &\quad
  + 2 H \qty(2 H^2 \dot{\phi}
  + 2 \dot{H} \dot{\phi}
  + H \ddot{\phi} ) G_{5X}
  + H X \qty(4 H^2 \dot{\phi}
  + 4 \dot{H} \dot{\phi}
  + 10 H \ddot{\phi} ) G_{5XX}
  + 4 H^2 \ddot{\phi} X^2 G_{5XXX}.
\end{align}
We integrate out the nondynamical valuables and obtain the quadratic action written solely in terms of the dynamical variables.

\begin{figure}
  \begin{center}
    \includegraphics[keepaspectratio=true,width=0.49\textwidth]{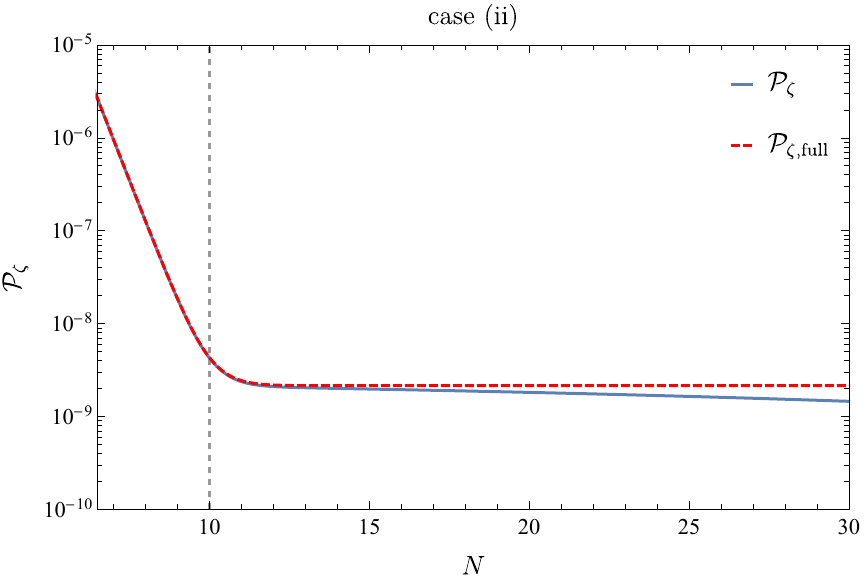}
    \includegraphics[keepaspectratio=true,width=0.49\textwidth]{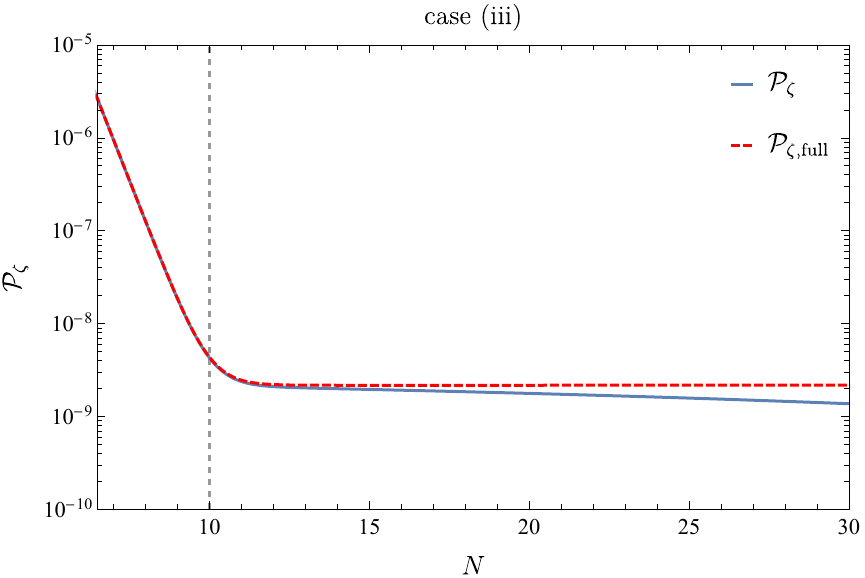}
  \end{center}
  \caption{Comparison of the evolution of $\calp_{\zeta}$ with and without metric perturbations in \ac{CMB} scale with $k=k_{\textrm{CMB}}$.
  Left and right panels are for the model (ii) and the model (iii), respectively.
  Blue solid lines show $\calp_{\zeta}$ obtained by ignoring metric perturbations,
  while red dashed lines represent $\calp_{\zeta}$ obatined by taking into account all the metric perturbations.
  Vertical dashed lines show the horizon crossing time. 
  }
  \label{fig: spt_Full_G5X2}
\end{figure}

Let us introduce the slow-roll parameters defined as
\begin{align}
  \epsilon = - \frac{\dot{H}}{H^2},
  \quad
  \epsilon_{\phi} = \frac{\dot{\phi}^2}{2\mpl^2 H^2},
  \quad
  \eta = \frac{\ddot{\phi}}{H\dot{\phi}},
  \quad
  \epsilon_{E} = \frac{\qty(\dot{Q}+HQ)^2}{\mpl^2 H^2},
  \quad
  \epsilon_{B} = \frac{\ga^2 Q^4}{\mpl^2 H^2}.
\end{align}
In the present setup, these quantities are small during inflation.
We may assume $\dot{Q}=0$ because the gauge field sits at the minimum of the effective potential.
We then replace $(\dot{H},\dot{\phi},\ddot{\phi},Q,\ga)$
in the equations of motion with the slow-roll parameters
$(\epsilon,\epsilon_{\phi},\eta,\epsilon_{E},\epsilon_{B})$,
keeping terms of first order in the slow-roll parameters.
We solve the resultant equations of motion numerically
and compare the power spectrum of the curvature perturbation, $\calp_{\zeta,{\rm full}}$,
with that obtained by ignoring the metric perturbations from the beginning, $\calp_\zeta$.
Figure~\ref{fig: spt_Full_G5X2} shows the comparison of the power spectra computed in the two different ways.
Here, we consider the models (ii) (left) and (iii) (right).
(From our result for the model (ii), one can confirm the replication of the result of Ref.~\cite{Dimastrogiovanni:2023oid}.)
We see that the curvature perturbation is conserved on superhorizon scales when the metric perturbations are taken into account.
After horizon crossing, we see only small deviations between the two spectra, implying that the approximation of ignoring the metric perturbations is good.
In particular, by evaluating the power spectra soon after horizon crossing, one can obtain a sufficiently accurate result.
Let $N_{\rm obs}$ be the time when we evaluate the power spectra.
In this paper, we choose $N_{\rm obs} = N_{\rm HC} + \Delta N$, where $N_{\rm HC}$ is horizon crossing time and
$\Delta N= 2$. 
This is the same value as used in Ref.~\cite{Dimastrogiovanni:2023oid}.

\section{Quantization of a coupled system} \label{App: Quantaization}
Following Ref.~\cite{Nilles:2001fg}, we briefly summarize the quantization method for a coupled system.
First, we will introduce the general procedure and
then rewrite the results into formulas that can easily be applied to the specific system we are interested in.
In this appendix, we use the conformal time $\tau$.

The general procedure for quantizing a coupled system is as follows.
We consider a set of dynamical variables $v_{I}$, where $I=1,2,\dots, N$ is a label of each component.
In general, the kinetic terms for $v_{I}$ are not of the canonical form,
but we can always perform a rotation in field space, $\Delta_{I}=\calr_{IJ} v_{J}$, to obtain canonical kinetic terms.
Then, the action for $\Delta_I$ is written in the form
\begin{align}
  S = \frac{1}{2} \int \dd{\tau} \dd[3]{x}
  \qty[\Delta'_{I} \Delta'_{I}
  + \Delta'_{I} K_{IJ} \Delta_{J}
  - \Delta_{I} K_{IJ} \Delta'_{J}
  - \Delta_{I} \Omega_{IJ}^{2} \Delta_{J}],
  \label{action: cannonical}
\end{align}
where $K_{IJ}$ is an anti-symmetric matrix and $\Omega^2_{IJ}$ is a symmetric matrix. We start with the action written in this form.
The equation of motion that follows from the above action is given by
\begin{align}
  \Delta''_{I} + 2 K_{IJ} \Delta'_{J}
  + \qty(K'_{IJ} + \Omega_{IJ}^{2} ) \Delta_{J}
  =0.
  \label{eq: EOMdelta}
\end{align}
We can eliminate the terms with the coefficients $K_{IJ}$ by further performing a rotation in field space, $\psi_{I} = \calq_{IJ} \Delta_{J}$,
where $\calq_{IJ}$ is an orthogonal matrix that satisfies $\calq'_{IJ}=\calq_{IL} K_{LJ}$.%
\footnote{An orthogonal matrix $\calq$ satisfies $\calq^{\top} \calq = \calq \calq^{\top} = I,$ where $\top$ is a transpose and $I$ is an identity matrix.
Differentiating this relation with repsect to time, we obtain $\calq^{\top} \calq' =- (\calq^{\top} \calq')^{\top}$, which shows that
$\calq^{\top} \calq'$ is an anti-symmetric matrix.
Choosing appropriately the $N(N-1)/2$ independent components of $\calq$, one can find $\calq$ satisfying $\calq^{\top} \calq'=K$ for any anti-symmetric matrix $K$.}
Using the matrix $\calq_{IJ}$, we can write the action as
\begin{align}
  S = \int \dd{\tau} \dd[3]{x} {\cal L}
  = \frac{1}{2} \int \dd{\tau} \dd[3]{x}
  \qty(\psi'_{I} \psi'_{I} - \psi_{I} {\cal S}^2_{IJ} \psi_{J}),
  \quad
  {\cal S}^2_{IJ}
  = \calq_{IL} \qty(\Omega^{2}_{LM} - K_{LN} K_{NM}) \calq_{MJ}^{\top},
\end{align}
where $\top$ denotes the transpose of the matrix.
The conjugate momentum is defined as
\begin{align}
  \Psi_{I} := \pdv{\cal L}{\psi'_{I}} = \psi'_{I},
\end{align}
and the Hamiltonian reads
\begin{align}
  H = \frac{1}{2} \int \dd[3]{x}
  \qty(\Psi_{I} \Psi_{I} + \psi_{I} {\cal S}^2_{IJ} \psi_{J}).
  \label{eq; Hamiltonian}
\end{align}
In general, the matrix ${\cal S}^2_{IJ}$ is nondiagonal.

One then proceeds to the usual quantization procedure for each set of the canonical variables $\Psi_{I}$ and $\psi_{I}$.
We perform the decomposition
\begin{align}
  \hat{\psi}_{I}
  = C_{IJ}\int \frac{\dd[3]{k}}{(2\pi)^3}
  \qty(u_{JK} \hat{a}_{K} + u^{*}_{JK} \hat{a}^{\dagger}_{K} )
  e^{-i\bm{k}\cdot\bm{x}},
  \quad
  \hat{\Psi}_{I}
  =C_{IJ} \int \frac{\dd[3]{k}}{(2\pi)^3}
  \qty(\tilde{u}_{JK} \hat{a}_{K} + \tilde{u}^{*}_{JK} \hat{a}^{\dagger}_{K} )
  e^{-i\bm{k}\cdot\bm{x}},
  \label{eq: ModeExpansion}
\end{align}
where $C_{IJ}$, an orthogonal matrix diagonalizing $\mathcal{S}_{IJ}^2$:
\begin{align}
  C_{IK}^{\top} {\cal S}^2_{KL} C_{LJ}
  = \diag \qty(\omega^{2}_{(1)},...,\omega^{2}_{(N)})
  =: \omega^2_{IJ},
\end{align}
is introduced for later convenience.
We impose the commutation relations
\begin{align}
  \qty[\hat{\psi}_{I} (\bm{x}), \ \hat{\Psi}_{J} (\bm{y})  ]
  = i \delta^3 (\bm{x} - \bm{y}) \delta_{IJ},\quad \text{others} = 0,
  \label{eq: Quantize}
\end{align}
for the conjugate fields
and 
\begin{align}
  \qty[\hat{a}_{I} (\bm{k}), \ \hat{a}_{J}^\dagger (\bm{p})  ]
  = (2\pi)^3 \delta^3 (\bm{k} - \bm{p}) \delta_{IJ},
  \quad \text{others} = 0,
  \label{eq: CAoperator}
\end{align}
for the creation and annihilation operators $\hat{a}_{I}$ and $\hat{a}^{\dagger}_{I}$.
Equations~\eqref{eq: Quantize} and~\eqref{eq: CAoperator} are satisfied if
\begin{align}
  u_{IK} \tilde{u}_{JK}^{*} - u_{IK}^{*} \tilde{u}_{JK}
  = i \delta_{IJ}.
  \label{eq: norm_u}
\end{align}
The canonical equations of motion are given by
\begin{align}
  \psi'_{I} = \Psi_{I},
  \quad
  \Psi'_{I} = - {\cal S}^2_{IJ} \psi_{J},
\end{align}
which yield
\begin{align}
  u'_{IJ} = \tilde{u}_{IJ} - \Gamma_{IK} u_{KJ},
  \quad
  \tilde{u}'_{IJ}
  =- \Gamma_{IK} \tilde{u}_{KJ}
  - \omega^2_{IK} u_{KJ},
  \label{eq: EOMmode}
\end{align}
where $\Gamma_{IJ}=C^{\top}_{IK}C'_{KJ}$ is anti-symmetric.
Now, $N$ second-order differential equations~\eqref{eq: EOMdelta} for $\Delta_{I}$ have reduced to $N\times N$ differential equations~\eqref{eq: EOMmode} for $u_{IJ}$.
While the $N$ diagonal components of the Wronskian condition~\eqref{eq: norm_u} are used to determine the normalization of the fields,
the remaining $N^2-N$ components give constraints among $u_{IJ}$,
We are thus left with $N^2-(N^2-N)=N$ independent components of $u_{IJ}$.
The number of dynamical degrees of freedom remains unchanged.

Next, let us discuss the initial adiabatic vacuum of this system, specializing to the case of inflationary cosmological perturbations with multiple degrees of freedom.
Now, we assume that the background field is constant in the early stage of the evolution of the system.
In the sub-horizon limit, we have $\Omega^2_{IJ}\simeq {\cal O}(k^2) \delta_{IJ}\gg K_{IK}K_{KJ}$ in the specific cases we are considering, as can be seen from the explicit expressions for these matrices (see Appendix~\ref{App: Perturbation}).
It then follows that $\Delta_I\simeq \psi_I$ and $\mathcal{S}_{IJ}^2\simeq\Omega_{IJ}^2=\mathcal{O}(k^2)\delta_{IJ}$ in the early time.
Since $\mathcal{S}^2_{IJ}$ is already diagonal in this limit, we have $C _{IJ}\simeq \delta_{IJ}$, and hence $\Gamma_{IJ}\simeq 0$, leading to
\begin{align}
  u'_{IJ} \simeq \tilde{u}_{IJ},
  \quad
  \tilde{u}'_{IJ} \simeq - \omega^2_{IK} u_{KJ}
  \quad\Rightarrow\quad
  u''_{IJ} \simeq - \omega^2_{IK} u_{KJ}.
\end{align}
The solution is given by
\begin{align}
  u_{IJ}
  = C_{1} \exp \qty[-i \int^{\tau} \omega_{(I)} \D \tilde{\tau}] \delta_{IJ}
  + C_{2} \exp \qty[i \int^{\tau} \omega_{(I)} \D \tilde{\tau}] \delta_{IJ},
\end{align}
where $C_{1}$ and $C_{2}$ are integration constants.
We consider the adiabatic vacuum condition and hence set $C_2=0$.
Using the normalization condition~\eqref{eq: norm_u},
we obtain the mode function at some initial moment as
\begin{align}
  u_{IJ} = \frac{1}{\sqrt{2\omega_{(I)}}} \delta_{IJ},
  \quad
  \tilde{u}_{IJ} = -i \sqrt{\frac{\omega_{(I)}}{2}}  \delta_{IJ},
\end{align}
where we omitted irrelevant phase factors.

Having thus determined the initial conditions for $u_{IJ}$,
we can solve the equations of motion and accordingly find the time evolution of the original variables $v_I$ by multiplying the (inverse of the) matrices $C_{IJ}$, $\calq_{IJ}$, and $\mathcal{R}_{IJ}$.
We therefore need to specify the form of these matrices explicitly.
These steps allow for some simplification, as shown in Refs.~\cite{Gumrukcuoglu:2010yc, Dimastrogiovanni:2012ew, Dimastrogiovanni:2023oid}
and as presented below.

Practically, it is more convenient to deal with $\Delta_I$ directly by performing the decomposition
\begin{align}
  \hat{\Delta}_{I} = \int \frac{\dd[3]{k}}{(2\pi)^3}
  \qty[\cald_{IJ}(\tau,k) \hat{a}_{J}(\bm{k})
  + \cald^{*}_{IJ}(\tau,k) \hat{a}^{\dagger}_{J}(\bm{k})]
  e^{-i \bm{k}\cdot\bm{x}},
\end{align}
where $\cald_{IJ}$ is related to $u_{IJ}$ as
\begin{align}
  \cald_{IJ} = \calq^{\top}_{IK} C_{KL} u_{LJ},
  \quad
  \cald'_{IJ}
  = - K_{IK} \calq^{\top}_{KL} C_{LM} u_{MJ}
  + \calq^{\top}_{IK} C_{KL} \tilde{u}_{LJ}.
  \label{eq: FieldReDefinition}
\end{align}
The equation of motion is given by
\begin{align}
  \cald''_{IL} + 2K_{IJ} \cald'_{JL}
  + \qty(K'_{IJ} + \Omega^2_{IJ} ) \cald_{JL} = 0.
  \label{eq: EOMforPert}
\end{align}
In the sub-horizon limit, the frequency matrix takes the form
\begin{align}
  \Omega^2_{IJ} \simeq \omega^2_{IJ} = k^2 c^2_{IJ},
\end{align}
where $c^2_{IJ}$ is a diagonal matrix whose $I$-th element corresponds to the propagation speed of the $I$-th field.
In this limit, the appropriate initial conditions are given by
\begin{align}
  \cald_{IJ} = \frac{1}{\sqrt{2k}} c_{IJ}^{-1/2},
  \quad
  \cald'_{IJ} = -i \sqrt{\frac{k}{2}} c_{IJ}^{1/2},
\end{align}
where we assume that $c^2_{IJ}$ is a slowly varying function of time in the early time.
In the case of the conventional \ac{CNI} model, we have $c_{IJ}^{2} = \delta_{IJ}$.
However, the propagation speeds of the scalar modes may differ from unity in the Horndeski family of scalar-tensor theories.
In terms of $\mathcal{D}_{IJ}$, the power spectrum is evaluated as
\begin{align}
  \calp_{IJ} (k) = \frac{k^3}{2\pi^2}
  \calr_{IL} \cald_{LK} \cald_{KM}^{*} \calr_{MJ}^{*}.
  \label{eq: PowerSpectrum}
\end{align}

To summarize, in this Appendix we have introduced the quantization method for a coupled system.
One can quantize the system by identifying the matrices $\calr_{IJ}, K_{IJ}, \Omega^{2}_{IJ}$, and $c^2_{IJ}$ in the action of our system.
In practice, $c^2_{IJ}$ can be read off from $\Omega^{2}_{IJ}$.

\section{Matrices appearing in the quadratic actions for scalar and tensor perturbations} \label{App: Perturbation}

From the quadratic actions for scalar and tensor perturbations,
one can identify the explicit form of the matrices introduced in Appendix~\ref{App: Quantaization}. For the tensor perturbations, we have
\begin{align}
  \calr_{t,IJ} &= 
  \begin{pmatrix}
    \displaystyle{\frac{2}{a\sqrt{\calg_{T}}}} &0
    \\
    0 &\displaystyle{\frac{1}{a}}
  \end{pmatrix},
  \quad
  K_{t,IJ} =
  \begin{pmatrix}
    0 &\displaystyle{\frac{a}{\sqrt{\calg_{T}}} \qty(\dot{Q} + HQ )}
    \\
    \displaystyle{\frac{a}{\sqrt{\calg_{T}}} \qty(\dot{Q} + HQ )} &0
  \end{pmatrix},
  \notag \\
  \frac{\Omega_{t,11}^2}{a^2}
  &= \frac{\calf_{T}}{\calg_{T}} \frac{k^2}{a^2}
  - \dot{H} - 2 H^2
  - \frac{2}{\calg_{T}} \qty[\qty(\dot{Q} + H Q  )^2 - \ga^2 Q^4 ]
  - \frac{\ddot{\calg}_{T}}{2 \calg_{T}}
  - \frac{3 H \dot{\calg}_{T}}{2 \calg_{T}}
  + \frac{\dot{\calg}_{T}^2}{4 \calg_{T}^2},
  \notag \\
  \frac{\Omega_{t,12}^2}{a^2} 
  &= \frac{\Omega_{t,21}^2}{a^2}
  = \frac{1}{\sqrt{\calg_{T}}} \qty[
    2 \rho_{\pm} \ga Q^2 \frac{k}{a}
  + \qty(H + \frac{\dot{\calg}_{T}}{2 \calg_{T}} )
  \qty(\dot{Q} + H Q  )
  - \frac{\ga \lambda}{f} Q^2 \dot{\phi}
   ],
  \notag \\
  \frac{\Omega_{t,22}^2}{a^2} &=
  \frac{k^2}{a^2}
  - \rho_{\pm} \frac{k}{a} \qty(2 \ga Q + \frac{\lambda}{f} \dot{\phi} )
  + \frac{\ga \lambda}{f} Q \dot{\phi},
  \notag \\
  c^{2}_{t,IJ} &= 
  \begin{pmatrix}
    \displaystyle{\frac{\calf_{T}}{\calg_{T}}} &0
    \\
    0 &1
  \end{pmatrix},
\end{align}
where $\rho_{\pm}=\pm 1$ for each polarization mode,
\begin{align}
  {\cal F}_{T} =
  2\qty(G_{4} - \ddot{\phi} X G_{5X}),
\end{align}
and ${\cal G}_{T}$ was already given in Eq.~\eqref{eq: calgT}:
$\mathcal{G}_T=2\left(G_4-2XG_{4X}-HX\dot\phi G_{5X}\right)$.

For the scalar perturbations, the matrices are given by
\begin{align}
  \calr_{s,IJ} &= 
  \begin{pmatrix}
    \displaystyle{\frac{2}{a\sqrt{\cala_{S}}}} &0 &0
    \\
    0 &\displaystyle{\frac{1}{\sqrt{2} a}} &0
    \\
    0 &\displaystyle{\frac{1}{\sqrt{2} k^2 a}}
    &\displaystyle{\frac{\sqrt{k^2 + 2\ga^2 a^2 Q^2}}{\sqrt{2} k^2 \ga a^2 Q}}
  \end{pmatrix},
  \notag \\
  K_{s,IJ} &=
  \begin{pmatrix}
    0 & \displaystyle{\frac{\ga \lambda  a Q^2}{f \sqrt{2\cala_{S}}}}
    & \displaystyle{-\frac{\ga^2 \lambda  a^2 Q^3}{f \sqrt{2\cala_{S}}
    \sqrt{k^2 + 2 \ga^2 a^2 Q^2}}}
    \\
    \displaystyle{-\frac{\ga \lambda  a Q^2}{f \sqrt{2\cala_{S}}}} & 0 & 0
    \\
    \displaystyle{\frac{\ga^2 \lambda  a^2 Q^3}{f \sqrt{2\cala_{S}} \sqrt{k^2 + 2 \ga^2 a^2 Q^2}}} & 0 & 0 \\
  \end{pmatrix},
\notag \\ 
  \frac{\Omega_{s,11}^2}{a^2}
  &=
  \frac{\calb_{S}}{\cala_{S}} \frac{k^2}{a^2}
  + \frac{\mathcal{C}_{S}}{\cala_{S}}
  - \dot{H}
  - 2 H^2
  - \frac{\ddot{\cala}_{S}}{2 \cala_{S}}
  - \frac{3 H \dot{\cala}_{S}}{2 \cala_{S}}
  + \frac{\dot{\cala}_{S}^2}{4 \cala_{S}^2}
  + \frac{\ga^2 \lambda ^2 k^2 Q^4}{f^2 \cala_{S} \left(k^2 + 2 \ga^2 a^2 Q^2\right)}
  ,
  \notag \\
  \frac{\Omega_{s,12}^2}{a^2} 
  &= \frac{\Omega_{s,21}^2}{a^2}
  = \frac{\ga \lambda Q}{f \sqrt{2 \cala_{S}}} \qty[
  2 \dot{Q} + 3 H Q
  + \frac{Q}{2} \frac{\dot{\cala}_{S}}{\cala_{S}}
  ],
  \notag \\
  \frac{\Omega_{s,13}^2}{a^2} 
  &= \frac{\Omega_{s,31}^2}{a^2}
  \notag\\
  &= - \frac{\sqrt{2} \lambda }{f \sqrt{\cala_{S}}}
  \qty[
  \frac{\ga^2 a H Q^3}{2 \sqrt{k^2 + 2 \ga^2 a^2 Q^2}}
  + \frac{2 k^4 + 3 \ga^2 k^2 a^2 Q^2 + 4 \ga^4 a^4 Q^4}{2 a \left(k^2 + 2 \ga^2 a^2 Q^2\right)^{3/2}} \left(\dot{Q} + H Q\right)
  + \frac{\ga^2 a Q^3}{4 \sqrt{k^2 + 2 \ga^2 a^2 Q^2}}
  \frac{\dot{\cala}_{S}}{\cala_{S}}
  ],
  \notag \\
  \frac{\Omega_{s,22}^2}{a^2}
  &= \frac{k^2}{a^2} + 4 \ga^2 Q^2 - \ga Q \frac{\lambda \dot{\phi}}{f},
  \notag \\
  \frac{\Omega_{s,23}^2}{a^2} &= \frac{\Omega_{s,32}^2}{a^2}
  = - \frac{\sqrt{k^2 + 2 \ga^2 a^2 Q^2}}{a}
  \qty(2 \ga Q - \frac{\lambda \dot{\phi}}{f} )
  ,
  \notag \\
  \frac{\Omega_{s,33}^2}{a^2}
  &= \frac{k^2}{a^2}
  + \frac{4 \ga^2 Q^2 \left(k^2 + \ga^2 a^2 Q^2\right)}{k^2 + 2 \ga^2 a^2 Q^2}
  - \frac{\lambda \dot{\phi}}{f} \frac{\ga k^2 Q }{k^2 + 2 \ga^2 a^2 Q^2}
  + \frac{6 \ga^2 k^2 a^2 \left(\dot{Q} + H Q\right)^2}{\left(k^2 + 2 \ga^2 a^2 Q^2\right)^2},
  \notag \\
  c^{2}_{s,11} &= \frac{\calb_{S}}{\cala_{S}},
  \quad
  c^{2}_{s,22} = c^{2}_{33} = 1,
\end{align}
where
$\cala_S$, $\calb_S$, and $\mathcal{C}_S$
were already defined in Appendix~\ref{app: MetricPtb}.

\section{On the curvature perturbation \texorpdfstring{$\zeta$}{zeta}}\label{App: CurvaturePtb}

In the main text, the curvature perturbation on uniform density slices, $\zeta$, is approximated by Eq.~\eqref{eq: defCurvaturePtb} on superhorizon scales.
In this appendix, we discuss the validity of this approximation.

The curvature perturbation on uniform density slices is written as
\begin{align}
  \zeta = \frac{\delta\rho}{3(\rho + p)}
  =\frac{\rho_\phi}{\rho+p}\cdot \frac{\delta\rho_\phi}{3\rho_\phi}
  +\frac{\rho_Q}{\rho+p}\cdot\frac{\delta\rho_Q}{3\rho_Q},
\end{align}
where $\rho=\rho_\phi+\rho_Q$ is the background energy density, $p=p_\phi+p_Q$ is the background pressure, and $\delta\rho=\delta\rho_\phi+\delta\rho_Q$ is the perturbation of the total energy density in the spatially flat gauge, with $\delta\rho_\phi$ and $\delta\rho_Q$ being the scalar-field and the gauge-field pieces, respectively.
In the present case, we have $\rho_\phi\gg \rho_Q$, and hence
$\zeta\simeq \delta\rho_\phi/3(\rho+p)$.
In Appendix~\ref{app: MetricPtb} we show that the metric perturbations can be ignored under the slow-roll approximation.
Under the same approximation, $\delta \rho_\phi$ is given by 
\begin{align}
  \delta\rho_\phi
  &= - \qty(G_{2\phi}
  - \dot{\phi}^2 G_{2\phi X}) \dphi
  + \dot{\phi} \qty(G_{2X}
  + \dot{\phi}^2 G_{2XX}) \ddphi
  + \dot{\phi} \qty(
   9 H \dot{\phi} G_{3X}
  + 3 H \dot{\phi}^3 G_{3XX}) \ddphi
  \notag\\
  &\quad
  + 6 H \qty(
   3 H \dot{\phi} G_{4X}
  + 6 H \dot{\phi}^3 G_{4XX}
  + H \dot{\phi}^5 G_{4XXX}) \ddphi
  + H^2 \dot{\phi} \qty(
   15 H \dot{\phi} G_{5X}
  + 10 H \dot{\phi}^3 G_{5XX}
  + H \dot{\phi}^5 G_{5XXX}) \ddphi
  \notag\\
  &\quad
  + {\cal O}(k^2),
\end{align}
where $\mathcal{O}(k^2)$ terms are ignored on superhorizon scales.
In our inflation models, we have 
\begin{align}
    \delta\rho_\phi \simeq V_\phi\delta\phi +\mathcal{O}(H\cali\dot\phi)\times
    \frac{\ddphi}{H}.
\end{align}
We have $\delta\phi\gg \ddphi/H$ on superhorizon scales, as confirmed by numerical solutions, and hence the curvature perturbation on uniform density slices is approximately given by
\begin{align}
    \zeta\simeq \frac{V_\phi\delta\phi}{3(\rho+p)}\simeq -\frac{H}{\dot\phi}\delta\phi,
\end{align}
where we used the background equations with the slow-roll approximation.

\bibliography{refs}
\bibliographystyle{JHEP}

\end{document}